\documentclass[a4paper,11pt]{article}
\pdfoutput=1
\usepackage{amsmath}
\usepackage{amssymb}
\usepackage{color}
\usepackage{cite}
\usepackage{hyperref}
\usepackage{graphicx}
\usepackage{subfig}

\makeatletter
\@addtoreset{equation}{section}
\renewcommand{\theequation}{\thesection.\@arabic\c@equation}
\makeatother
\makeatletter
\renewcommand\appendix{\par%\newpage
  \setcounter{section}{0}%
  \setcounter{subsection}{0}%
  \gdef\thesection{Appendix \@Alph\c@section }
  \renewcommand{\theequation}
  {\Alph{section}.\arabic{equation}}
}
\makeatother
\def \be {\begin{equation}}
\def \ee {\end{equation}}
\def \ba {\begin{array}}
\def \ea {\end{array}}
\def \bea{\begin{eqnarray}}
\def \eea{\end{eqnarray}}

\def \b {\beta}

\def \f {\frac}

\def \nn {\nonumber}

\def \hs {\hspace}

\def \inf {\infty}

\def \Tr {{\textrm{Tr}}}

\title{\textbf{Holographic Calculation for Large Interval R\'enyi Entropy at High Temperature}}
\author{
Bin Chen$^{1,2,3}$\footnote{bchen01@pku.edu.cn}\,
and
Jie-qiang Wu$^{1}$\footnote{jieqiangwu@pku.edu.cn}
}
\date{}

\begin{document}

\maketitle

\begin{center}
{\it
$^{1}$Department of Physics and State Key Laboratory of Nuclear Physics and Technology, Peking University, Beijing 100871, P.R.\! China\\
\vspace{2mm}
$^{2}$Collaborative Innovation Center of Quantum Matter,  \\Beijing 100871, P.~R.~China\\
$^{3}$Center for High Energy Physics, Peking University,  \\Beijing 100871, P.~R.~China
}
\vspace{10mm}
\end{center}

\begin{abstract}
In this paper, we study the holographic R\'enyi entropy of a large interval on a circle at high temperature for the two-dimensional  CFT  dual to pure AdS$_3$ gravity. In the field theory,  the R\'enyi entropy is encoded in the  CFT partition function on $n$-sheeted torus connected with each other by a large branch cut. As proposed in \cite{Chen:2014hta}, the effective way to read the entropy in the large interval limit is to insert a complete set of state bases of the twist sector at the branch cut. Then the calculation transforms into an expansion of four-point functions in the twist sector with respect to $e^{-\frac{2\pi TR}{n}}$.  By using the operator product expansion of the twist operators at the branch points, we read the first few terms of the R\'enyi entropy, including the leading and next-to-leading contributions in the large central charge limit. Moreover, we show that the leading  contribution is actually captured by the twist vacuum module. In this case by the Ward identity the four-point functions can be derived from the correlation function of four twist operators, which is related to double interval entanglement entropy.
Holographically,  we apply the recipe in \cite{Faulkner:2013yia} and \cite{Barrella:2013wja} to compute the classical R\'enyi entropy and its 1-loop quantum correction, after imposing a new set of monodromy conditions.  The holographic classical result matches exactly with the leading contribution in the field theory  up to $e^{-4\pi TR}$ and $l^6$, while the holographical 1-loop contribution is in exact agreement  with next-to-leading results in field theory up to $e^{-\frac{6\pi TR}{n}}$ and $l^4$ as well. %Our results support $AdS_3/CFT_2$ correspondence in some degree.

\end{abstract}

\baselineskip 18pt
\thispagestyle{empty}

\newpage

\section{Introduction}

The entanglement entropy is an important notion in a quantum many-body system \cite{nielsen2010quantum,petz2008quantum}. Not only could it be used to measure the effective degrees of freedom in the system, but it could also be taken as a quantum order parameter, among its various applications. It is defined as follows. Let $A$ be a subsystem, and then the reduced density matrix of $A$ is obtained by tracing out the degrees of freedom of its complement$A^c$
\be \rho_A=\Tr_{A^c} \rho, \ee
where $\rho$ is the density matrix of the whole system. Then the entanglement entropy is defined to be the von Neumann entropy of the reduced density matrix
\be\label{SEE} S_{A}=-\Tr \rho_A \log \rho_A. \ee
Furthermore for pure state $\rho=\mid \psi \rangle \langle \psi \mid$,
the entanglement entropy of the subsystem is equal to the one of its complementary part
\be S_A=S_{A^{c}}, \ee
but for a thermal state
%\be \rho=\frac{e^{-\beta H}}{Z}, \ee
the equality breaks down,
\be S_A \neq S_{A^c} \ee
because of the thermal effect.
  It is convenient to calculate the entanglement entropy from the R\'enyi entropy, which is defined to be
\be S_n=-\frac{1}{n-1} \log\Tr \rho_A^n.\ee
The entanglement entropy can be read from
\be S_{EE}=\lim_{n\rightarrow 1}S_n, \ee
if the limit $n \rightarrow 1$ is well defined. %(However, it is not easy to take this limit, as shown
% in \cite{Calabrese:2009ez} and see also \cite{DeNobili:2015dla} for some numerical calculation to this problem. )% and in this paper.

In quantum field theory,  the entanglement entropy and R\'enyi entropy are hard to compute because there are  an infinite number of degrees of freedom. In this case, the entanglement entropy is
defined with respect to a spatial submanifold at a fixed time.  By using the replica trick\cite{Callan:1994py} the R\'enyi entropy can be transformed into the partition function of $n$ copies of field theory with the fields being identified at the submanifold. It is usually a formidable task to compute this partition function for a general field theory. Even for two-dimensional (2D) conformal field theory (CFT), which is expected to give more analytic results due to the existence of infinite dimensional symmetries, the exact results are limited. For 2D CFT, the R\'enyi entropy is generally related to the partition function on a higher genus Riemann surface. Besides a few universal results determined by the conformal  symmetries\cite{Calabrese:2004eu}, only the partition functions of a free boson and fermion on a higher genus Riemann surface have been known \cite{Calabrese:2009ez} \cite{Herzog:2013py} \cite{Chen:2014hta} \cite{Datta:2013hba}\cite{Headrick:2012fk} \cite{Lokhande:2015zma}.

However, it is possible to expand the partition function with respect to some modular parameters for a general CFT in some cases. The two simplest nontrivial examples are the case of two intervals on a complex plane and  the case of one interval on a torus. %the resulting Riemann surface has only two different kinds of cycles, $A$ cycle and $B$ cycle, others being related to them by the replica symmetry.
Generally, to calculate the partition function, one can cut the Riemann surface at some cycles and insert a complete set of state bases  such that the full Riemann surface changes into a surface without  handle and  hole and the computation transforms into a summation of multipoint correlation functions on a full complex plane. The key point is to find the nice way to cut open the Riemann surface such that the expansion series is well behaved.
For a general genus-$g$ Riemann surface, we can always choose $g$ couples of $A$ cycles and $B$ cycles with a proper intersection \cite{Dijkgraaf:1987vp} and cut the Riemann surface at certain cycles.
Different choices on the cutting  correspond to different ways of expanding the partition function. Even though by the modular invariance the different expansions should be equal to each other, their convergent rates are different.

The simplest trivial example is the partition function on a torus.  One may quantize the theory along the thermal direction or the spatial direction, which corresponds to inserting  the complete bases along the spatial cycle or thermal cycle, and the partition function could be written as
\be Z=\sum_i e^{-\frac{2\pi \beta}{R}(L_i+\bar{L}_i-\frac{c}{12})},\hs{3ex}\mbox{or}\hs{2ex}Z
=\sum_i e^{-\frac{2\pi R}{\beta}(L_i+\bar{L}_i-\frac{c}{12})}. \ee
Because of the modular invariance, the two different calculations  give the same answer. At a  low temperature the  quantization along the thermal direction leads to a better convergent series  while at a high temperature the spatial quantization works better.

%CFT$_n$

\begin{figure}
  \centering
  % Requires \usepackage{graphicx}
  \includegraphics[width=5cm]{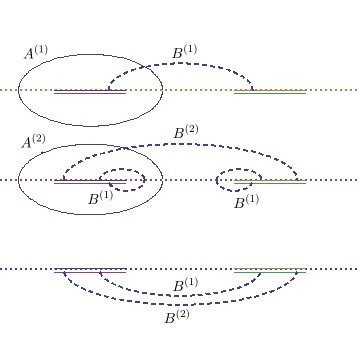}\\
  \caption{Canonical cycles for the double intervals}\label{cycle1}
\end{figure}

For the double-interval case, if the intervals are short, one may  take the operator product expansion(OPE) of the twist operators to compute the R\'enyi and entanglement entropy order by order  with respect to a small cross ratio\cite{Headrick:2010zt,Calabrese:2010he}.
Actually, taking OPE is equivalent to inserting a complete set of bases at the cycles around the two intervals on every sheet. In the Riemann surface for the $n$th R\'enyi entropy of the double-interval case, there are $n-1$ independent couples of cycles denoted by $A_i$ and $B_i$. As shown in Fig. \ref{cycle1} for the $n=3$ case, there are two couples of independent cycles $A_i,B_i, i=1,2$. In the small interval limit,
 we can take the OPE of the  twist operators at the branch points of the first interval. This is equivalent to cutting and inserting a complete set of state bases at $A_i$ cycles enclosing the interval. This expansion is well convergent for a small cross ratio.
On the contrary,  for a large cross ratio which means the intervals are large and the branch points of two intervals are close to each other,  we must take the OPE of the second and third twist operators, which is equivalent to cutting the Riemann surface along the $B_i$ cycles enclosing the branch points of separated intervals. %\footnote{We show the cycles for double interval in Appendix D as \ref{cycle1}.}

For the case of a single interval on a torus,  the Riemann surface is obtained by connecting $n$ tori  along the branch cut. When the interval is not very large, one may cut the spatial or thermal cycle and insert a complete set of bases to compute the partition function\cite{Herzog:2013py} \cite{Chen:2014unl}, just as in the genus-1 case. Which cycles to cut depends on the temperature. But for a high temperature and a very large interval, the previous treatment is not good enough because the resulting expansion series is poorly behaved. Instead, it was proposed in \cite{Chen:2014ehg} \cite{Chen:2014hta} that one should cut the cycle crossing the branch cut. This requires the insertion of a complete set of bases in the twist sector rather than the normal sector in the $n$-copied CFT. This proposal has been checked for the free compact and noncompact bosons, and it has been applied to prove the universal relation between the thermal entropy and the entanglement entropy.

The AdS/CFT correspondence provides another way to compute the entanglement entropy in a CFT. For the Einstein gravity, it was first proposed by Ryu and Takayanagi\cite{Ryu:2006bv,Ryu:2006ef} that the entanglement entropy could be holographically given by the area of a minimal surface in the bulk, which is homogeneous to $A$
 \be S_{EE}= \frac{\mbox{Area}(\Sigma_A)}{4G}. \ee
 The holographic entanglement entropy could be understood as a generalized gravitational entropy \cite{Lewkowycz:2013nqa}  \cite{Fursaev:2006ih}. In the higher dimension case,  it is not clear if the holographic entanglement entropy gives precisely the entanglement entropy in the dual field theory. Nevertheless, for a 2D CFT holographically dual to AdS$_3$ gravity, it has been proved that the holographic computation is correct in the semiclassical regime\cite{Hartman:2013mia,Faulkner:2013yia}.  Therefore, the R\'enyi entropy provides a new window to study the AdS$_3$/CFT$_2$ correspondence.

The $AdS_3/CFT_2$ correspondence states that the quantum gravity in $AdS_3$ spacetime is dual to a 2D CFT with a central charge \cite{Brown:1986nw}
 \be c=\frac{3l}{2G}, \ee
and a sparse light spectrum \cite{Hartman:2013mia} \cite{Hartman:2014oaa}, where $G$ is the three-dimensional (3D) gravity coupling constant. Though a precise definition of AdS$_3$ quantum gravity, possibly a string theory,  has not been well established, its semiclassical limit has been much studied. As the classical configurations in the AdS$_3$ gravity could be obtained as the quotients of the global AdS$_3$ by the subgroup of the isometry group $SL(2,C)$, the path integral of semiclassical AdS$_3$ gravity could be defined in principle. On the other side, the explicit construction of dual CFT is not known. Nevertheless, the large central charge limit of the CFT, corresponding to the semiclassical gravity, is much simplified. Under this limit, only the vacuum module dominates the contribution to the CFT partition function\cite{Hartman:2013mia}. As a result, the partition function is universal in the sense that it is very much restricted by the conformal symmetry, and is independent of the explicit construction of the CFT.  In this work, we are interested in the large central charge limit of the R\'enyi entropy of 2D CFT. From the AdS$_3$/CFT$_2$ correspondence, the  partition function of the Riemann surface in the CFT should  be given by  the partition function of the 
gravitational configuration ending on the Riemann surface. In the large central charge limit, 
   the R\'enyi entropy can be decomposed into the terms proportional to $c$, $c^0$, $\f{1}{c}$, ...,  which should correspond respectively to   the classical, quantum 1-loop, 2-loop, etc. , parts of the gravitational partition function\cite{Headrick:2010zt}.

In the field theory side, for a genus-$g$ Riemann surface, we need to choose $g$ $A_i$ cycles to insert  complete bases at each cycle such that the expansion converges fast. %For the CFT which dual to a pure gravity, we only insert a vacuum module at these cycles.
Under the large $c$ limit, the dominant contribution to the partition function comes from the light primary states \cite{Hartman:2013mia} and their descendants. The heavy states  give only nonperturbative corrections of order $O(e^{-c})$. Furthermore,  among the light spectrums only the vacuum module gives the linear  $c$ order result and the other modules  give only higher order corrections with respect to $\frac{1}{c}$\cite{Hartman:2013mia}. %\footnote{In the original paper \cite{Hartman:2013mia}, they used different conformal block to expand the four twist operators' correlation function, and argue only the vacuum module give the leading large $c$ result. The light primary operators give perturbative correction and the heavy operators give non-perturbative correction. However, as we explained the original description equals our "cutting and inserting" description.}
Moreover, it turns out that even for the next-to-leading correction with respect to $\frac{1}{c}$, the first few terms in the expansion  are captured only by the vacuum module \cite{Maloney:2007ud} \cite{Chen:2013kpa} \cite{Chen:2014unl}.
The vacuum Verma module consists of a primary identity operator and its descendants which could be constructed by the stress tensors $T(z)$ and $\bar T(\bar z)$. In this work, we assume that for the field theory that is dual to the pure gravity, we only need to insert the vacuum module at each cycle.

%(Note that the vacuum module here means the one for the original CFT, not the one for the orbifold CFT or $n$-sheeted CFT. In other words, the states inserted at the cut cycles is the vacuum module of the original CFT in that sheet.  For example in the low temperature single interval case, the vacuum module in each sheet is inserted in the expansion of the thermal density matrix, and for the double interval case the operator product expansion include the operators constructed from the stress tensors in each replica.)

Because of the replica symmetry, we always deal with a CFT on an $n$-sheeted surface, which can be regarded as one CFT with $n$ copies of the original field with the fields in different replicas being identified along the branch cut. When we combine the $n$-sheeted surface's field into one CFT, we call it $n$-copied CFT\footnote{In the literature, this $n$-copied CFT is usually called orbifold CFT. As in our following discussion we often use the tensor product of $n$ copies of CFT, and we would like to call it the $n$-copied CFT. }  with $nc$ central charge and denote it as CFT$_n$. The original CFT is denoted as CFT$_1$.  If we do not consider the monodromy condition of the fields around the branch point, the $n$ copies of the fields are decoupled so that we have just a tensor product of $n$ copies of the fields. In this case, we call it the normal sector of CFT$_n$. In contrast, if we consider the twist monodromy condition of the fields around the branch point, we get the twist sector of CFT$_n$. In both cases,
 we can classify the states by the irreducible representations of its Virasoro algebra Vir$^{(t)}$, defined by the stress tensor $T(z)=\displaystyle{\sum_{j=1}^{n}T^{(j)}(z)}$.  Under Vir$^{(t)}$, the twist (normal) sector states can be decomposed into  more than one irreducible module, and the one with the lowest conformal dimension is called the twist (trivial) vacuum module. Note that the twist (trivial) vacuum module has a different meaning from the vacuum module in the original CFT.
For the partition function of a CFT on an $n$-sheeted Riemann surface resulting from the replica trick,   we have two pictures to compute it. One is to regard it as the one of  CFT$_1$ on the $n$-sheeted surface,  and the other one is to regard it as the correlator of twist operators in CFT$_n$. %We will transform between these two pictures in the calculation for convenience. When we insert or classify states, we will always point out which picture we use.

On the bulk side, to calculate the partition function on a higher genus Riemann surface holographically, one needs to find the gravity configuration whose asymptotic boundary is exactly the Riemann surface.  To find the gravity configuration, one can use the Schottky uniformization to get the Riemann surface and then extend the uniformization to the bulk.  However for one Riemann surface, there may be more than one Schottky uniformization and different uniformizations give different gravity configurations. Among all of the gravity configurations the one with the least classical action dominates  the partition function in the large $c$ limit\cite{Faulkner:2013yia}. The contributions from other configurations are suppressed as $O(e^{-c})$. Furthermore the 1-loop correction can be determined by the functional determinant of the fluctuations  around the classical background \cite{Barrella:2013wja} by using the heat kernel method developed in \cite{Yin:2007gv,Giombi:2008vd}.  Recently, by using the operator product expansion of the twist operators, the holographic computation of the double-interval R\'enyi entanglement entropy for the CFT  has been checked beyond the classical level\cite{Chen:2013kpa,Chen:2013dxa,Perlmutter:2013paa,Chen:2014kja,Beccaria:2014lqa}. Furthermore, for the single interval on a circle at finite temperature, if the interval is not very large, the holographic computation has been
confirmed to be in exact agreement with the field theory computation\cite{Chen:2014unl}, in which  the thermal density matrix is expanded level by level \cite{Cardy:2014jwa}.

In this paper, we study the  R\'enyi entropy of a large interval on a circle at high temperature in the context of AdS$_3$/CFT$_2$ correspondence, extending our previous study
in \cite{Chen:2014unl}. In the large interval limit, the computations in both the field theory and the bulk need to be developed furthermore. On the field theory side, the study in \cite{Chen:2014unl} showed that the perturbative series in the partition function do not converge well. Actually, the classical part of the R\'enyi entropy is just
\bea\label{class} S_n\mid_{classical}
&=&\frac{c}{6}\frac{1+n}{n}\log\sinh(2\pi TY)+\mbox{const.}-\left.\frac{c}{9}\frac{(n+1)(n^2-1)}{n^3}\right\{\sinh^4(2\pi Ty)e^{-4\pi TR} \notag\\
&~&+4\sinh^4(2\pi Ty)\cosh^2(2\pi Ty)e^{-6\pi TR}+\left(\frac{-11-2n^2+1309n^4}{11520n^4}\cosh(16\pi Ty) \right.\notag \\
&~&-\frac{-11+28n^2+199n^4}{1440n^4}\cosh(12\pi Ty)-\frac{77-346n^2+197n^4}{2880n^4}\cosh(8\pi Ty) \notag \\
&~&\left.\left.-\frac{-77+436n^2+433n^4}{1440n^4}\cosh(4\pi Ty)+\frac{-77+466n^2+907n^4}{2304n^4}\right)e^{-8\pi TR} \right\} \notag \\
&~&+O(e^{-10\pi TR}), \eea
where $y$ is the length of the interval. When the  length of the interval is comparable with the size of the circle $y\sim R$, the expansion converges very slowly and is not good anymore. This asks us to find another perturbative way to compute the partition function more effectively and reliably. In \cite{Chen:2014hta}, we proposed to  insert a complete twist sector states through the branch cut and expand the R\'enyi entropy with respect to
$e^{-\frac{2\pi L}{n \beta}}$.  In \cite{Chen:2014ehg,Chen:2014hta}, we tested this proposal and reconsidered the noncompact and compact free scalars
and found good agreements with direct expansions of the partition functions. Now we are going to consider the CFT with a holographic gravity dual,
in the large interval and high temperature limit. We only consider the vacuum module of CFT$_1$ and its correspondents in the twist sector of CFT$_n$.  After cutting through the branch cut, the Riemann surface still have a nonzero genus and the four-point functions in the twist sector cannot be calculated directly. Furthermore, we use the OPE of the two twist operators at the branch points and compute the correlation functions on the unfolded cylinder of length $n\beta$ with the fields in the OPE at the different positions. We manage to  expand the result with respect to complementary part of the interval length $l$. %\footnote{This correspond to cut the Riemann surface in the $a_i$ cycles in \ref{cycle2}.}
We calculate the R\'enyi entropy up to order $l^4$ and $e^{-\frac{8\pi L}{nR}}$, including the leading linear $c$, the $c$-independent, and $1/c$ parts. Moreover, we find that the leading contribution is actually captured by the twist vacuum module. We support this result not only by the argument using the  large central charge limit of the conformal blocks but also by direct computation using the Ward identity. As a result, we obtain the exact formula for the entanglement entropy.%,  as suggested in \cite{Azeyanagi:2007bj}.

%On the other hand, by wald identity, we also calculate the contribution from twist vacuum's decedents generated by total Virasoro algebra up to $e^{-\frac{4\pi}{\beta}}$ and $l^6$. And we argue that the linear c term for R\'enyi entropy can be fixed by these calculations.

\begin{figure}
  \centering
  % Requires \usepackage{graphicx}
  \includegraphics[width=5cm]{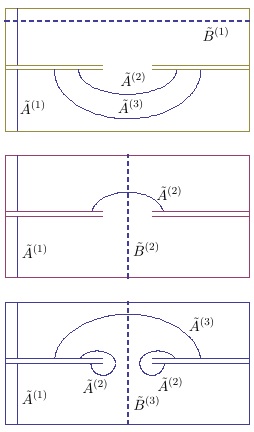}\\
  \caption{Canonical cycles for the large interval at high temperature: $n=3$ case}\label{cycle2}
\end{figure}

For the holographic calculation, we follow the treatments in  \cite{Barrella:2013wja} \cite{Faulkner:2013yia}, but basing on a different  monodromy condition. As shown in \cite{Azeyanagi:2007bj}, the holographic entanglement entropy for the large interval case is not read from the bulk geodesic ending on the interval. Instead, it is the sum of the horizon length and the geodesic of the complementary interval. This fact suggests that there is a phase transition when the interval becomes large, and the bulk gravitational configuration for the large interval must be different.  Instead of the cycles used in \cite{Chen:2014unl}, we choose another $n$ cycles to be of  trivial monodromy. Among them, there is one cycle crossing the branch cut $n$ times, and the other $n-1$ independent ones crossing the branch cut and enclosing the complementary part of the branch cut in different sheets. As shown in Fig. \ref{cycle2} for $n=3$, we set $\tilde A_i$'s to be of trivial monodromy. As a warm-up, we compute the classical holographic entanglement entropy by using the new monodromy conditions, and we obtain the result suggested in \cite{Azeyanagi:2007bj}. Furthermore, we compute the holographic R\'enyi entropy up to $l^6$ and $e^{-4\pi TR}$ for classical contribution, and up to $l^4$ and $e^{-\frac{6\pi TR}{n}}$ for 1-loop quantum contribution. The results are in perfect match with CFT's computation.

%We calculate the R\'enyi entropy For classical order we can read the entanglement entropy, which is the same as

The remaining parts of the paper are organized as follows. In Sec. 2, we present the field theory computation. After a brief review on the twist sector of the CFT$_n$,
we focus on the vacuum module and compute  the R\'enyi entropy in the first few orders.  In Sec. 3, we show how to do holographic computation with the new monodromy condition. We obtain both the classical and 1-loop quantum results perturbatively. Up to the orders  we are interested in, we find good agreements with the field theory results. In Sec. 4, we end with conclusions and discussions.
We collect some technical details in Appendices.

\section{Field theory calculation}

\begin{figure}[tbp]
\centering
\subfloat[$n$ sheeted Riemann surface]{\includegraphics[width=4cm]{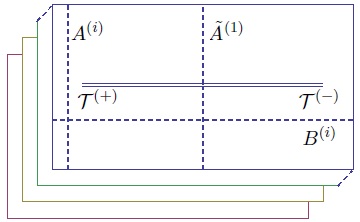}\label{i1}}
\quad
\subfloat[translate the interval]{\includegraphics[width=4cm]{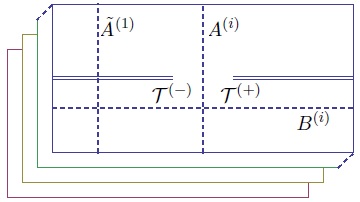}\label{i2}}
\quad
\subfloat[unfold the twist]{\includegraphics[width=4cm]{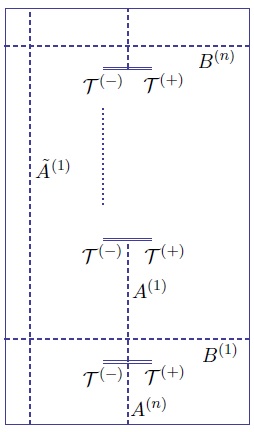}\label{i3}}
\\
\caption{Riemann surface for finite temperature R\'enyi entropy. The horizontal line denotes the spatial direction of unit length, and the vertical line denotes the thermal direction of length $\b$. (a) The Riemann surface for $n$ cylinders connected by a branch cut, which is denoted by a solid line ${\cal{T}}^{(+)}{\cal{T}}^{-}$. In each sheet, there is a cycle $A^{(i)}$ along the imaginary time direction. However there could also be a cycle marked by $\tilde{A}^{(1)}$, which crosses the branch cut and goes from one sheet to the next one $n$ times until it goes back to the original sheet.   (b) The same as (a), just by shifting the branch cut to the boundary. ${\cal{T}}^{-}{\cal{T}}^{+}$ denotes the complement of the original interval. In (c) we unfold the twist. There are $n$ copies of ${\cal{T}}^{-}{\cal{T}}^{+}$ cuts in the unfolded picture. }
\end{figure}

In this section, we present our computation on the large interval R\'enyi entropy on a circle at high temperature in the CFT which corresponds to pure $AdS_3$  gravity. By the replica trick the R\'enyi entropy can be transformed into calculating the partition function on a higher genus Riemann surface, which is obtained by pasting $n$ tori along a large branch cut. In Fig. 2, we cut open the torus and show the branch cut (interval). The horizontal line is the spatial direction of length $R$, and the vertical line is the thermal direction of a length $\beta$. The large interval  is presented as double solid lines between ${\cal{T}}^{(+)}$ and ${\cal{T}}^{(-)}$. In Fig. \ref{i2}, we translate the interval,  and in Fig. \ref{i3}, we unfold the branch cut and get a cylinder of thermal length $n\beta$, with $n$ cuts. We denote the coordinate in Fig. \ref{i2} as $u$ and the one in Fig. \ref{i3} as $y$, and set the two branch points to be  at $\pm \frac{l}{2}$ in Fig. \ref{i2}.

As in our previous paper \cite{Chen:2014hta} \cite{Chen:2014ehg}, we would cut the Riemann surface through the branch cut, $\tilde{A}^{(1)}$ cycle in Fig. \ref{i2}, and insert  complete bases in the twist sector of CFT$_n$ at the cycle. The calculation transforms into a series of four-point functions with two twist operators and two operators in the twist sector. As the two branch points are actually very close to each other, we may take the OPE for the two twist operators, which  amounts to an infinite summation of local operators in the $n$-copied   CFT.

Alternatively, we may unfold the Riemann surface as in Fig. \ref{i3} and insert  complete state bases of the normal sector in single sheet CFT to do the computation. Changing into the coordinate in Fig. \ref{i3},  we find that the localized operators  in the $j$th copy in the OPE sit in the imaginary axis
\be y^{(j)}=i(j\beta). \ee
In this way, each term in the expansion is a multipoint correlation function on an infinite cylinder, with two normal sector operators at the left and the right infinities of the cylinder\footnote{By the state-operator correspondence, the inserted normal sector states can be transformed into two vertex operators at the left and the right infinities with a factor $e^{-\frac{2\pi R}{n\beta}\Delta}$.} and $n$ localized  operators at $y^{(j)}$.  %As we assumed, in the inserting and OPE we only consider the vacuum module in each sheet of the theory.

%(However, we note that the vacuum module in each copy of CFT is different from the vacuum module in the $n$-copied CFT which is defined with respect to  the Virasoro algebra stemmed from the total stress tensor $T(z)=\sum_{j=1}^{n}T^{(j)}(z)$. This means that in the calculation we need to consider the states generated by the $n$ copies of Virasoro algebras (expanded from $n$ copies of stress tensors) rather than those generated by the total Virasoro algebra (expanded from total stress tensor). Later we will see that the latter one  captures the classical contribution correctly.)

We calculate the leading terms for the R\'enyi entropy expanded with respect to both $e^{-\frac{2\pi R}{n\beta}}$ and $l$. The results in the large $c$ limit actually include the leading  contribution, which is linear $c$,  and the next-to-leading contribution, which is of order one, and even the next-to-next-to-leading contribution. We argue that only the descendants of the  twist vacuum, the states generated by acting the  Virasoro algebra Vir$^{(t)}$ of CFT$_n$ on the twist vacuum, contribute to the leading $c$ result. We confirm this fact by the analysis of the classical conformal block expansion in the large $c$ limit. We furthermore derive the first few leading linear $c$ terms from the descendants by using the Ward identity on the correlation functions of four twist operators, which could be related to the one in the double-interval case. In the $n\to 1$ limit, we find that only the twist vacuum module has a nonvanishing contribution. This leads to  the entanglement entropy of the large interval, which is exactly the same as the one from the holographic computation.  However, when we consider the next-to-leading contribution to the R\'enyi entropy, we have to take  the contributions from other states in the twist sector into account. %We calculate the classical R\'enyi and entanglement entropy expanded with respect to $e^{-\frac{2\pi R}{\beta}}$ and $l$.

\subsection{Twist sector}

In this subsection, let us give a brief review on the twist sector in the CFT$_n$. By the replica trick the R\'enyi entropy can be transformed into the partition function of a single copy CFT on an $n$-sheeted surface connected at the branch cut.  From the path integral, it is easy to see that the partition function could be taken as the $n$ copies of field theory, one on each sheet, with fields on different sheets being related at the branch cut. In this $n$-copied theory, the locality requires us to introduce the twist field or antitwist field at the branch points\cite{Lunin:2000yv,Calabrese:2004eu}.  %The twist field is a primary field with conformal dimension  $\D_n= We can combine the field on  $n$-sheeted surface together as one CFT with $n$-copied original field, with twist monodromy condition at the branch point. For convenience, we will call the states in $n$-copied theory with no branch cut and no twist boundary condition as trivial sector states.

\begin{figure}
  \centering
  % Requires \usepackage{graphicx}
  \includegraphics[width=7cm]{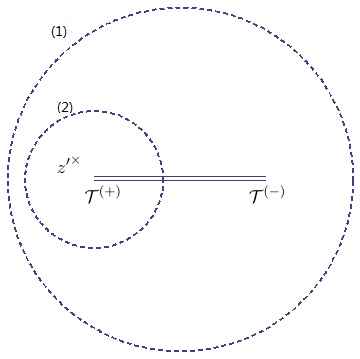}\\
  \caption{Branch cut}\label{branch}
\end{figure}

Let us show how the twist sector arises in a CFT$_n$ with a branch cut, following the discussion in \cite{Dixon:1986qv}.  As the twist field is a local field, we consider simply the  $n$-sheeted surface connected by a single branch cut.   As in Fig.\ref{branch}, the ${\cal{T}^{\pm}}$ denote the branch point located at $z_1,z_2$ and the double line denotes the branch cut. Now we study it as a CFT$_n$. Considering an operator $T^{(j_1)}(z)$ , it will change to $T^{(j_1+1)}(z)$ when it moves once around the branch point ${\cal{T}^+}$ along the circle (2). The point $z=z_1$ is a branch singularity in the $n$-copied theory, which is a source of stress tensor \cite{Calabrese:2004eu}
\be <T(z)>=n<T^{(1)}(z)>\sim\frac{c}{24}(n-\frac{1}{n})\frac{1}{(z-z_1)^2}, \ee
when $z$ is close to $z_1$. Independence on the branch cut implies  a local operator at the branch point. This local operator is known as the twist operator, denoted as ${\cal{\hat T}^{+}}(z_1)$. It is a primary field, with conformal dimension $h=\frac{c}{24}(n-\frac{1}{n})$. In a similar way, we can get the antitwist operators at the other branch point. Much information of the CFT$_n$ is encoded in the twist operators. For example, the partition function of an $n$-sheeted complex plane with $N$ intervals is determined by the $2N$-point function of the twist and antitwist operators on a complex plane. Moreover, considering the operator-state correspondence, the twist operator corresponds to the ground state in the twisted sector of Hilbert space. Considering the OPE of the twist field with other basic fields in the theory, we will find other excited twist fields. Correspondingly, we find the excited states in the twist sector, as we will review soon.

%The property of the twist operator could be read from the operator product expansion of it with other fields. not only encodes the universal information of the CFT$_n$ with branch cut, but also the monodromy condition of the fields. Considering the OPE of the twist operator with the stress tensor, the resulting operators include the ones

%As we see the twist operator will introduce non-singularity in the $n$-copied theory, which makes it different from trivial sector states. Furthermore, if we set another operator for example $T^{(j_2)}(z^{'})$ close to the twist operator, by operator product expansion, there will be new operators appearing. Moving a test operator $T^{(j_1)}(z)$ around circle (2), it will still change into $T{(j_1+1)}(z)$, which means the new operators from OPE are still out of trivial sector. Actually all of the operators generated in this way will introduce non-singularity which build the twist sector. In the same way at the point $z_2$, we have the antitwist operator and can generate the antitwist sector.

Before our discussion of the twist sector and antitwist sector states in the $n$-copied field theory, we show that the OPE of a twist sector operator and an antitwist sector operator give trivial sector operators in the $n$-copied field theory. As argued above, the excited states in the twist sector could be obtained by considering the monodromy of the field moving around the branch point $z_1$ along the circle (2) in  Fig.\ref{branch} . However, when we consider the OPE of the operators in both the twist sector and antitwist sector, by the monodromy condition, the resulting states in the circle (1) in  Fig.\ref{branch} must be in the trivial sector. This fact has been applied in the discussion of the OPE of two twist operators in the short interval limit. In that case, the operators in the expansion are in the tensor product of the normal sector of the CFT in each sheet (which is just the trivial sector in the CFT$_n$ picture), as shown in \cite{Calabrese:2010he,Chen:2013kpa}. More generally, for the excited states in the twist sector, their OPE should consist of the trivial sector states.

%Even though moving around the branch point and $z_2$ will have non-singularity effect, moving around both of the two branch point along circle (1) is always singularity, which means the states in circle (1) is always in trivial sector.
%We note that, here the twist sector is only to discuss partition function on the Riemann surface with branch cut, and we don't need relative locality for all these operators including trivial sector and twist sector.

Now, let us give a  review on the twist sector states \cite{Chen:2014unl} in the CFT$_n$ arising from the replica trick in calculating the R\'enyi entropy. Let us work in the coordinate $u$ in Fig. \ref{i2}. To expand the partition function, we need to insert  complete twist sector bases along the cycle $\tilde{A}^{(1)}$. We may temporarily forget about the geometric structure of the torus and only consider the geometry and the monodromy condition near the cycles $\tilde{A}^{(1)}$ . Moreover, as the vacuum module dominates in the large central charge limit, we focus on the twist sector from the vacuum module.  For more complete discussion on other modules,  please see \cite{Chen:2014unl} for more details. In the vacuum module, the fields are constructed from the stress tensor. The monodromy condition on the stress tensor is
\be T^{(j)}(u+i\beta)=T^{(j+1)}(u), \hspace{3ex}j=0,\cdots, n-1\ee
in the $u$ coordinate, with $ T^{(n)}(u)=T^{(0)}(u)$,
 and
\be T(y+in\beta)=T(y), \ee
in the $y$ coordinate.
 In the $y$ coordinate, the $n$ copies of fields are  unfolded  as
\be T^{(j)}(u)=T(y)\mid_{y=u+ij\beta}. \ee
Taking the conformal transformations
\be z=e^{\frac{2\pi u}{\beta}}, \ee
\be w=e^{\frac{2\pi y}{n \beta}}, \ee
the monodromies in the new coordinates $z,w$ are, respectively, 
\be T^{(j)}(ze^{2\pi i})=T^{(j+1)}(z), \ee
\be T(we^{2\pi i})=T(w). \ee
The states inserted at the $\tilde{A}^{(1)}$ cycle in Figs. \ref{i2} and Fig. \ref{i3} can be described as the vertex operators being inserted at the origin of $z$ and $w$. As in \cite{Chen:2014hta}, we can redefine the operators in the $z$ coordinate,
\be\label{newT} T^{(t,k)}(z)\equiv\sum_{j=1}^{n}T^{(j)}(z)e^{\frac{2\pi i}{n}kj}, \hs{3ex}k=0,1,\cdots,n-1 \ee
and expand it as
\be T^{(t,k)}(z)=\sum_{m\in {\cal Z}}\frac{L^{(k)}_m}{z^{m+2+\frac{k}{n}}}. \ee
The operators $L^{(k)}_m$ satisfy a commutation relation similar to the Virasoro algebra. Among the operators $T^{(t,k)}(z)$,  $T^{(t,0)}$ is of special importance. It is the total stress tensor for the whole $n$-copied theory, and $\{L^{(0)}_m\}$ are the generators of the corresponding  Virasoro algebra Vir$^{(t)}$.

We may study the spectrum of the theory with respect to Vir$^{(t)}$. From the commutators between $L^{(k)}_m$ and  $L_0^{(0)}$,
\be [L^{(0)}_0,L^{(k)}_m]=-(m+\frac{k}{n})L^{k}_m. \ee
we know that when  the operators $L_m^{(k)}$ act on a state, those with
\be ~ m>0, ~0\leq k<n, ~~
 \mbox{or}~~~  m=0, ~0<k<n, \ee
decrease  conformal dimension, so they are annihilation operators; while those with
\be ~m<0, ~0\leq k<n, \ee
increase the  conformal dimension, so they are creation operators. Therefore we can define the vacuum for the twist sector to be
\bea L^{(k)}_m\mid t\rangle=0 ~~~\mbox{for}&~& m>0, ~0\leq k<n, ~~
 \mbox{or}~~~  m=0, ~0<k<n. \eea
The twist vacuum  has the lowest conformal dimension
\be
h_v=\frac{c}{24}n(1-\frac{1}{n^2}).
\ee
Acting with the creation operators on the twist vacuum we can get all of the excited states in the twist sector.

There is a one-to-one correspondence between the twist sector states in the CFT$_n$ and the normal sector states in the original one-sheet CFT. Actually  the trivial monodromy condition in the $w$ coordinate suggests that the mode expansion in the $w$ coordinate for the field gives the normal sector of the CFT. The conformal dimensions between the twist sector states and the normal sector states are related by
\be\label{tw} h_{twist}=\frac{1}{n}h_{normal}+\frac{c}{24}n(1-\frac{1}{n^2}). \ee
On the cylinder,  the energy of the state could be  written as
\bea\label{en} H&=&\frac{2\pi}{\beta}(L_{twist}+\bar{L}_{twist}-\frac{nc}{12})
=\frac{2\pi}{n\beta}(L_{normal}+\bar{L}_{normal}-\frac{c}{12}). \eea
The $\frac{1}{n}$ factor  in the last equation is due to the fact that in the $y$ coordinate  the length of the thermal cycle is $n\beta$. For convenience, we will denote the states in the twist sector as $\mid t,i \rangle$, corresponding to the state $\mid i\rangle $ in  the original theory.

It turns out to be more useful to classify the states in the twist sector by using the  conformal symmetry in the $n$-copied theory. %First notice that the $T^{(t,0)}(z)$ is the stress tensor for the $n$-copied theory and has trivial monodromy. Its mode expansion define a Virasoro algebra with a central charge $nc$.
The states should be decomposed into different irreducible modules of the Virasoro algebra Vir$^{(t)}$. To show this decomposition we calculate the chiral partition function for the twist sector
\be\label{twist} \Tr q^{L_0}=q^{h_v}\prod_{s=2}^{\inf}\frac{1}{1-q^{\frac{s}{n}}}= \big( q^{h_v}\cdot\prod_{
s\geq 2, \frac{s}{n} \notin \mathbb{N}}
\frac{1}{1-q^{\frac{s}{n}}} \big)
\cdot\prod_{r=1}^{\inf}\frac{1}{1-q^r} .\ee
 In the second equation, we have used the fact that there is a one-to-one correspondence between the twist sector states in the CFT$_n$ and the normal sector states in the one-sheet CFT, with their the conformal dimensions being  related by (\ref{tw}). For the vacuum module in the normal sector, the descendants are generated by the Virasoro algebra $L_{-s}$ with $s\geq 2$, so the product begins from $s=2$.
For a primary operator with the conformal dimension $h$, if there is no null state in its descendants, its contribution to the chiral partition function is
\be \Tr q^{L_0}\mid_{h}=q^{h}\prod_{s=1}^{\inf}\frac{1}{1-q^{s}}. \ee
Considering this fact, the quantity in the parentheses of (\ref{twist}) can be taken as a generating function for the primary operators with respect to the Virasoro algebra. As explained in \cite{Chen:2014hta}, the operators $L^{(0)}_{-m}$ in the $z$ coordinate  correspond to the generators $L_{-mn}$  in the unfolded $w$ coordinate with $mn>1$. As $L_{-mn}$ do not generate null states in the normal sector vacuum module,  the operators $L^{(0)}_{-m}$ do not generate null states in the twist sector either. Therefore there is no null state in the descendants of each primary states in the twist sector.
Expanding the function
\be \big( q^{h_v}\cdot\prod_{
s\geq 2, \frac{s}{n} \notin \mathbb{N}}
\frac{1}{1-q^{\frac{s}{n}}} \big), \ee
with respect to $q$, the coefficient before $q^h$ is the number of the primary operator with conformal dimension $h$. It is clear that in the twist sector there are many new primary states and the number of the primary states increases exponentially  with the conformal dimension. For example, acting with the operators $L^{(n-i)}_{-1}$  on the twist vacuum, the resulting states that have conformal dimensions  $h=h_v+\frac{i}{n}$
 are the primary states, since they can be annihilated  by the  operators $L_m^{(0)}, m>0$ in Vir$^{(t)}$.  %Even though the twist sector states in the orbifold CFT always have an one-to-one correspondence with normal sector states, the states corresponding to the same state in the normal sector, can be quiet different for different n.
Among the modules in the twist sector, the vacuum module generated by $L_{m}^{(0)}, m<0$ on the twist vacuum is the most important one in our following discussion. We call this module the twist vacuum module.

In the following discussion, we will meet  another notion, the normal sector, in the $n$-copied field theory.  It is defined with respect to the $n$-copied field theory without a branch cut, or the tensor product of the $n$ copies of Hilbert space of the normal sector of a single CFT. Moreover, as we only focus on the vacuum module of the CFT, we call the tensor product of $n$-copied vacuum module as the trivial sector in the $n$-copied field theory. Note that the stress tensor $T^{(t,0)}$ is still a well-defined quantity in the $n$-copied field theory without the branch cut. Therefore we may classify the states in the trivial sector by the Virasoro algebra Vir$^{(t)}$. In this case, we find that there are
 exponentially increasing primary operators with respect to Vir$^{(t)}$ in the trivial sector of the CFT$_n$ as well. Considering the chiral partition function, we have
\bea \Tr q^{L_0}&=&(\prod_{s=2}^{\inf} \frac{1}{1-q^s})^n \notag \\
&=&\prod_{s=2}^{\inf}\frac{1}{1-q^s}
+(1-q)\left(\prod_{r=2}^{\inf}\frac{1}{(1-q^r)^{n-1}}-1\right)\prod_{s=1}^{\inf}\frac{1}{1-q^s}. \eea
In the last equation, we decompose the whole partition function into the contribution from  different modules with respect to Vir$^{(t)}$. Each module is generated by  $L_{-m}^{(0)},m>0$ acting on the highest weight state. The first term denotes the module generated from the vacuum state, with zero conformal dimension, so the product starts from $s=2$. For the other primary operators, there are no null states in their descendants.  This is because 
 the primary state has nonzero conformal dimension, and the states $L_{-1}^{m} \mid h\rangle$ have nonzero norm. Considering $[L_{r},L_{-r}]=\frac{c}{12}r(r^2-1)+2rL_0$, for $r>1$, the commutator has a linear $c$ term. In the large c limit, all of the states have nonzero norm. To read the number of other primary states,
 we just need to expand the quantity
\be (1-q)\left(\prod_{r=2}^{\inf}\frac{1}{(1-q^r)^{n-1}}-1\right), \ee
with respect to $q$, such that the coefficient before $q^h$ is just the number of the primary states with a conformal dimension $h$. We can easily see  that the number of the primary states increases exponentially with  their conformal weights. In the following, we call the module generated from the vacuum a trivial vacuum module, with respect to Vir$^{(t)}$. %the trivial here denote the monodromy condition for the field.

%^To be clear, we distinguish the three definition, vacuum module, twist vacuum module and trivial vacuum module. The vacuum module denote the states in one sheet theory. When we deal with one sheet surface the states are always in vacuum module, for example the states inserted in cycle $\tilde{A}^{(1)}$ in coordinate \ref{i3} or the states located at each sheet of the two twist operators expansion.
%On the other hand, because of replica symmetry, we may also deal with n-sheets theory. For convenience, we can regard it as one field theory with n copies of field with twist monodromy condition or trivial monodromy condition. They are the twist sector or trivial sector of $n$-sheets CFT, which appears in the inserting states at $\tilde{A}^{(1)}$ cycle and the two twist operators' OPE in coordinate \ref{i2}. In both of the two condition, the total stress tensor $T(z)=\sum_{j=0}^{n-1}T^{(j)}(z)$ are always single valued. Expanding the total stress tensor, we get the total Virasoro algebra which control the symmetry of n copies field theory. The states can be classified by the total Virasoro algebra. The module with lowest primary conformal dimension are the twist vacuum module and trivial vacuum module.

\subsection{R\'enyi entropy}

As discussed previously, we can expand the partition function by inserting  complete twist sector bases in Fig. \ref{i2}  or complete normal sector bases in Fig. \ref{i3}. The partition function can be expanded as
\bea\label{expand} Z_n&=&\sum_i e^{-\frac{2\pi R}{\beta}(h_{t,i}+\bar{h}_{t,i}-n\frac{c}{12})}\langle t,i\mid
{\cal{\hat T}}^-(u_1){\cal{\hat T}}^+(u_2)\mid t,i\rangle\mid_u  \notag \\
&=&\sum_i e^{-\frac{2\pi R}{n\beta}(h_i+\bar{h}_i-\frac{c}{12})}\langle t,i\mid
{\cal{\hat T}}^-(u_1){\cal{\hat T}}^+(u_2)\mid t,i\rangle\mid_u,
\eea
in the $u$ coordinate, and in the second line we use (\ref{tw}). And the thermal partition function reads
\bea Z_1=\sum_i e^{-\frac{2\pi R}{\beta}(h_i+\bar{h}_i-\frac{c}{12})}. \eea

In the  large $c$ limit, we only need to consider the vacuum module, which  captures the perturbative effect. We list the first few states in the vacuum module and their vertex operators at the origin and the infinity in Appendix A. Such states give the first few leading order contributions to the R\'enyi entropy.

Each term in (\ref{expand}) is a four-point function with two twist operators at the branch points and two operators in the twist sector at the left and the right infinities.
\footnote{The bases we insert when cutting the Riemann surface are in the Schr\"odinger picture. In (\ref{expand}), we change them into the Heisenberg picture, with a factor $e^{-\frac{2\pi R}{\beta}H}$, and the states now correspond to the vertex operators at the infinities.}
For the CFT dual to pure gravity, we do not know exactly the analytic form of this correlation function. Nevertheless, when the twist operators are very close,  we may take the OPE of two twist operators
\be\label{OPE} {\cal{\hat T}}^-(-\frac{l}{2}){\cal{\hat T}}^+(\frac{l}{2})\sim c_n l^{-\frac{c}{6}n(1-\frac{1}{n^2})}V(0)\bar{V}(0), \ee
where
\bea V(u)&=&1+\sum_{i}\frac{1}{12}(1-\frac{1}{n^2})l^2T^{(i)}(u)
+\sum_i\frac{1}{488}(1-\frac{1}{n^2})l^4\partial^2T^{(i)}(u)  \notag \\
&~&+\sum_i\frac{1}{288}(1-\frac{1}{n^2})^2l^4(:T^{(i)}(u)^2:-\frac{3}{10}\partial^2T^{(i)}(u)) \notag \\
&~&+\sum_{i_1<i_2}[\frac{1}{8c}\frac{1}{n^4}\frac{1}{\sin^4\frac{\pi}{n}(i_1-i_2)}+\frac{1}{144}(1-\frac{1}{n^2})^2]
l^4T^{(i_1)}(u)T^{(i_2)}(u)+O(l^5), \eea
and the similar form for the antiholomorphic part $\bar V(u)$. In the operator product expansion, we only consider  the vacuum module in each sheet and ignore the other modules.

In practice, it is more convenient to unfold the twist and consider the correlation function in Fig. \ref{i3}. After transforming into the coordinate $y$, we find that  the operators in $V(0)$ are localized at
\be T^{(j)}(0) \rightarrow T(y)|_{y=ij\beta}, \ee
namely, the operators in different sheets are unfolded and located at different positions in the cylinder. %Now the partition function is actually a multi-point correlation function involving $n$ pairs of the twist operators  on a cylinder of length $n\beta$.
We may use the OPE of the twist operators to compute the partition function perturbatively.
Now the partition function can be expanded as a multipoint correlation function on the cylinder involving $n$ operators located at $y=ij\beta$, and two vertex operators at left and right infinities.
%We may use the OPE of the twist operators to compute the partition function perturbatively.
%Under such an expansion, each term is a multi-point correlation function on the cylinder.
We can furthermore take a conformal transformation into the $w$ coordinate, and do the calculation in a full complex plane. Formally, we still write
\bea Z_n&=&e^{\frac{\pi Rc}{6n\beta}}\langle t\mid {\cal{\hat T}}^-(u_1){\cal{\hat T}}^+(u_2)\mid t\rangle
(\sum_{i,\bar{i}}\frac{ \langle t,i,\bar{i}\mid {\cal{\hat T}}^-(u_1){\cal{\hat T}}^+(u_2)\mid t,i,\bar{i}\rangle}
{\langle t\mid {\cal{\hat T}}^-(u_1){\cal{\hat T}}^+(u_2)\mid t\rangle}
e^{-\frac{2\pi R}{n\beta}(h_i+\bar{h}_i)}) \notag \\
&=& e^{\frac{\pi Rc}{6n\beta}}c_nl^{-\frac{c}{6}n(1-\frac{1}{n^2})}
\langle t\mid V(0)\mid t \rangle \langle t\mid \bar{V}(0)\mid t\rangle
(\sum_i \frac {\langle t,i \mid V(0)\mid t,i \rangle}{\langle t\mid V(0) \mid t\rangle}
e^{-\frac{2\pi R}{n\beta}h_i})
(\sum_{\bar{i}} \frac{\langle t,\bar{i} \mid V(0)\mid t,\bar{i} \rangle}{\langle t\mid V(0) \mid t\rangle}
e^{-\frac{2\pi R}{n\beta}\bar{h}_i}) \notag  \\
&=& e^{\frac{\pi Rc}{6n\beta}}c_nl^{-\frac{c}{6}n(1-\frac{1}{n^2})}
(\langle t\mid V(0)\mid t \rangle
(\sum_i \frac {\langle t,i \mid V(0)\mid t,i \rangle}{\langle t\mid V(0) \mid t\rangle})
e^{-\frac{2\pi R}{n\beta}h_i})^2.
\eea
In the second line we take into  (\ref{OPE}), and the fact that the holomorphic and antiholomorphic parts decompose. In this case, the holomorphic and antiholomorphic parts are equal to each other. The R\'enyi entropy is
\bea\label{Sn} S_n&=&-\frac{1}{n-1}\log \frac{ Z_n}{Z_1^n} \notag \\
&=&-\frac{1}{n-1}((\log c_n-\frac{c}{6}n(1-\frac{1}{n^2})\log l+\frac{\pi Rc}{6n\beta}
+2\log \langle t\mid V(0)\mid t \rangle \notag \\
&~&+2\log (\sum_i\frac{\langle t,i\mid V(0)\mid t,i\rangle}{\langle t\mid V(0)\mid t\rangle}
e^{-\frac{2\pi R}{n\beta}h_i}))-
n(\frac{\pi Rc}{6\beta}+2\log(\sum_i e^{-\frac{2\pi R}{\beta}h_i}))).
\eea
Denote
\be f_i=\frac{\langle t,i\mid V(0)\mid t, i\rangle}{\langle t, 0\mid V(0)\mid t, 0\rangle}\left|_u
=\frac{\langle i\mid V(0)\mid i\rangle}{\langle 0\mid V(0)\mid 0\rangle}\right|_y, \ee
where in the last equation, we change into the $y$ coordinate. Taking $f_i$ into (\ref{Sn}), we get the first few terms of the $n$th R\'enyi entropy
\bea\label{Snt} S_n&=&-\frac{1}{n-1}\log c_n+\frac{c}{6}\frac{n+1}{n}\log l+\frac{\pi Lc}{6\beta}\frac{n+1}{n}
-\frac{2}{n-1}\log \langle t\mid V(0)\mid t \rangle  \notag \\
&~&-\frac{2}{n-1}(f_2e^{-\frac{4\pi R}{n\beta}}+f_3e^{-\frac{6\pi R}{n\beta}}+
(f_{4,1}+f_{4,2}-\frac{1}{2}f_2^2)e^{-\frac{8\pi R}{n\beta}} +O(e^{-\frac{10\pi R}{n\beta}})) \notag \\
&~&+\frac{2n}{n-1}(e^{-\frac{4\pi R}{\beta}}+e^{-\frac{6\pi R}{\beta}}+\frac{3}{2}e^{-\frac{8\pi R}{\beta}}
+O(e^{-\frac{10\pi R}{\beta}})).
\eea
The R\'enyi entropy is expanded with respect to $e^{-\frac{2\pi R}{\beta}}$ and $e^{-\frac{2\pi R}{n\beta}}$ . The two expansion arguments are not independent. In the above formula, we have actually done the computation in Fig. \ref{i3}. In other words, we have unfolded the twist and consider the insertion of the normal sector states
at the left and the right infinities of the cylinder. Meanwhile to calculate the analytic form of $f_i$, we also take the OPE of the twist operators which is an expansion with respect to the relative length of the two twist operators  Therefore, we have two kinds of expansion, one from the normal sector states and the other from the OPE of the twist operators. This results in two different expansions in $S_n$. %For the definition of the state $\mid 4,1\rangle$ and $\mid 4,2 \rangle$ see Appendix A.

The explicit expressions of $f_j$'s can be found in Appendix C.
 There are a few remarkable properties on $f_j$'s.
\begin{enumerate}
\item First of all, the $l^2$ terms in $f_j$'s  always take the form 
$\frac{j}{3}\frac{(-1+n^2)\pi^2}{\beta^2n^3}$. This is because the operator at $l^2$ order in the OPE is the stress tensor $T(z)=\sum_{j=1}^{n}T^{(j)}(z)$, whose correlation function  is fixed by the Ward identity.
\item Second, for each $f_j$ there exist some exceptional integers $n$ at which $f_j$ does not share a general formula and take specific form. This fact forbids an analytic continuation of the R\'enyi entropy to noninteger $n$ in order to calculate the entanglement entropy. It is because  in the OPE (\ref{OPE}) there is always a term like
\be \sum_{j_1<j_2}T^{(j_1)}T^{(j_2)}.\ee
The correlation function involving such a term includes the summation
\be \sum_{j_1-j_2=1}^{n-1}\sin^m(\frac{\pi (j_1-j_2)}{n}). \ee
When $m<0$, the summation has a universal formula for any integer $n$, while for $m>0$, it is more complicated. We can rewrite the sine function in the summations as
\be \sin^m(\frac{\pi (j_1-j_2)}{n})=\sum_{r}a_re^{\frac{2\pi ir(j_1-j_2)}{n}}. \ee
If we take a summation for each term we find
\be \sum_{j_1-j_2=1}^{n-1} e^{\frac{2\pi i r(j_1-j_2)}{n}}=
\left\{ \begin{array}{ccc}
-1, ~&~\mbox{for}~&\frac{r}{n} \notin \mathbb{N} \\
n-1,~&~\mbox{for}~&\frac{r}{n} \in \mathbb{N}
\end{array} \right.,
\ee
which shows the nonanalytic origin.
\item We notice that $f_j$ has no linear $c$ contribution for nearly all of $n$ but finite exceptions. Actually, when $\mid t,j\rangle$ is primary, there is no linear $c$ term in $f_j$. To understand this effect, we can first transform into a $z$ coordinate and expand the four point function by conformal blocks. Using the large $c$ conformal block \cite{Hartman:2013mia}, there should be no linear $c$ contribution. For nearly all of $n$ but finite exceptions, $\mid t,j\rangle$ is primary. For example, when $n$ is bigger than the conformal dimension of $\mid j\rangle$, which is the corresponding state in the normal sector, the state $\mid t,j\rangle$ is primary because it can be annihilated  by all of the Virasoro algebra generators $L_{m}^{(0)}$ with $m>0$.

\item Furthermore, we also notice that for the $n$ in which $f_j$ has no linear $c$ terms, $f_j$ always share a general formula up to $l^4$. As we discussed before, the non-analytic property comes from correlation functions
    \be\label{ana} \langle t,j\mid T^{(j_1)}(u)T^{(j_2)}(u) \mid t,j \rangle. \ee
    Because of the symmetry, the correlation function is a sum of
    \be\label{sumana} \sin^{m}\frac{\pi}{n}(j_1-j_2). \ee
     If there are some terms with $m>0$ the final result is non-analytic for $n$. However,  if $\mid t,j\rangle$ is primary, this cannot happen. Consider
    \bea\label{nonanalitic} \lefteqn{ \sum_{j_1<j_2} \langle t,j\mid T^{(j_1)}(u)T^{(j_2)}(u) \mid t,j \rangle}  \notag \\
    &=&\frac{n}{2}\sum_{j_1-j_2=1}^{n-1} \langle t,j\mid T^{(j_1)}(u)T^{(j_2)}(u) \mid t,j \rangle, \notag \\
    &=&\frac{n}{2}\sum_{j_1-j_2=1}^{n-1} \sum_{m} b_m\sin^m \frac{\pi}{n}(j_1-j_2).\eea
    On the other hand, we have
    \bea \sum_{j_1<j_2} T^{(j_1)}(u)T^{(j_2)}(u)&=&
    \frac{1}{2}(\sum_{j=0}^{n-1}T^{(j)}(u))^2-\frac{1}{2}\sum_{j=0}^{n-1}(T^{(j)}(u))^2 \notag \\
    &=&\frac{1}{2}T(u)^2-\frac{1}{2}\sum_{j=0}^{n}(T^{(j)}(u))^2. \eea
    For the primary state, the correlation function $\langle t,j\mid T(u)^2\mid t,j\rangle$ can be fixed by the Ward identity, and it is analytic for all $n$. The term $\langle t,j \mid (T^{(j)}(u))^2 \mid t,j \rangle$ does not depend on $j$, and it is also analytic, which means there is no $m>0$ terms in (\ref{nonanalitic}). It is not clear whether this property can be extended to a higher order of OPE expansion with respect to $l$.

\end{enumerate}

Here  we just list the first few leading order results of the R\'enyi entropy:
\bea S_2&=&-\log c_2+c\left(\frac{1}{4} \log l+\frac{\pi R}{4\beta}+
(\frac{\pi^2l^2}{96\beta^2}-\frac{61\pi^4l^4}{184320\beta^4})-\frac{\pi^4l^4}{64\beta^4}e^{-\frac{2\pi R}{\beta}}
-\frac{11\pi^4l^4}{64\beta^4}e^{-\frac{4\pi R}{\beta}}\right) \notag \\
&~&+\left\{-\frac{\pi^4l^4}{2048\beta^4}-(2+\frac{\pi^2l^2}{2\beta^2}+\frac{25\pi^4l^4}{384\beta^4})e^{-\frac{2\pi R}{\beta}}
-(2+\frac{3\pi^2l^2}{4\beta^2}+\frac{15\pi^4l^4}{256\beta^4})e^{-\frac{3\pi R}{\beta}}\right. \notag \\
&~&\left.+(1-\frac{3\pi^2l^2}{2\beta^2}-\frac{275\pi^4l^4}{384\beta^4})e^{-\frac{4\pi R}{\beta}}\right\}
+\frac{1}{c}\left(-\frac{3\pi^4l^4}{64\beta^4}e^{-\frac{2\pi R}{\beta}}
+\frac{\pi^4 l^4}{128\beta^4} e^{-\frac{3\pi R}{\beta}}-\frac{21\pi^4l^4}{64\beta^4}e^{-\frac{4\pi R}{\beta}}\right) \notag \\
&~&+O(l^5)+O(e^{-\frac{5\pi R}{\beta}})
\eea

\bea S_3&=&-\frac{1}{2}c_3+c\left(\frac{2}{9}\log l+\frac{2\pi R}{9\beta}+(\frac{\pi^2l^2}{243\beta^2}
-\frac{347\pi^4l^4}{1771470\beta^4})-\frac{128\pi^4l^4}{6561\beta^4}e^{-\frac{2\pi R}{\beta}}\right) \notag \\
&~&+\left\{-\frac{16\pi^4l^4}{177147\beta^4}
-(1+\frac{16\pi^2l^2}{81\beta^2}+\frac{800\pi^4l^4}{59049\beta^4})e^{-\frac{4\pi R}{3\beta}}
-(1+\frac{8\pi^2l^2}{27\beta^2}+\frac{280\pi^4l^4}{6561\beta^4})e^{-\frac{2\pi R}{\beta}} \right.\notag \\
&~&\left. -(\frac{3}{2}+\frac{16\pi^2l^2}{27\beta^2}+\frac{608\pi^4l^4}{6561\beta^4})e^{-\frac{8\pi R}{3\beta}}\right\}
+\frac{1}{c}\left(-\frac{256\pi^4l^4}{59049\beta^4}e^{-\frac{4\pi R}{3\beta}}
-\frac{32\pi^4l^4}{6561\beta^4}e^{-\frac{2\pi R}{\beta}}-\frac{64\pi^4l^4}{6561\beta^4}e^{-\frac{8\pi R}{3\beta}}\right)
\notag \\
&~&+O(l^4)+O(e^{-\frac{10\pi R}{3\beta}})
\eea

\bea S_4&=&-\frac{1}{3}c_4+c\left(\frac{5}{24}\log l+\frac{5\pi R}{24\beta}+
(\frac{5\pi^2l^2}{2304\beta^2}-\frac{2089\pi^4l^4}{14155776\beta^4})
-\frac{126\pi^4l^4}{6144\beta^4}e^{-\frac{2\pi R}{\beta}}\right) \notag \\
&~&+\left\{-\frac{11\pi^4l^4}{262144\beta^4}
-(\frac{2}{3}+\frac{5\pi^2l^2}{48\pi^2}+\frac{749\pi^4l^4}{147456\beta^4})e^{-\frac{\pi R}{\beta}}
-(\frac{2}{3}+\frac{5\pi^2l^2}{32\beta^2}+\frac{1375\pi^4l^4}{98304\beta^4})e^{-\frac{3\pi R}{2\beta}} \right.\notag \\
&~&\left. -(1+\frac{5\pi^2l^2}{16\beta^2}+\frac{719\pi^4l^4}{16384\beta^4})e^{-\frac{2\pi R}{\beta}}\right\}
+\frac{1}{c}\left(-\frac{19\pi^4l^4}{24576\beta^4}e^{-\frac{\pi R}{\beta}}
-\frac{95\pi^4l^4}{49152\beta^4}e^{-\frac{3\pi R}{2\beta}}
-\frac{23\pi^4l^4}{8192\beta^4}e^{-\frac{2\pi R}{\beta}}\right) \notag \\
&~&+O(l^4)+O(e^{-\frac{5\pi R}{2\beta}})
\eea

\bea S_n
&=&-\frac{1}{n-1}\log c_n+c\left(\frac{1}{6}\frac{n+1}{n} \log l +\frac{\pi R}{6\beta}\frac{n+1}{n}
+\frac{\pi ^2 l^2 (n+1)}{36 \beta^2 n^3}-\frac{l^4\pi^4 (n+1)(n^6+9n^4-9n^2+11)}{12960\beta^4 n^7}\right) \notag \\
&~&+-\frac{l^4 \pi^4(n+1)(n^2+11)(3n^4+10n^2+227)}{226800 \beta^4 n^7}
-\frac{2}{n-1}\left\{ \left(1+\frac{2\pi^2 l^2(n^2-1)}{3\beta^2 n^3} \right.\right.\notag \\
&~&\left.-\frac{\pi^4 l^4(n^6-12n^5-3n^4+24n^3+135n^2-444n+299)}{54\beta^4n^7}\right)e^{-\frac{4\pi R}{n\beta}} \notag \\
&~&+\left(1+\frac{\pi^2 l^2 (n^2-1)}{\beta^2 n^3}
-\frac{\pi^4 l^4 (n^6-18n^5-5n^4+36n^3+835n^2-4626n+6081)}{36 \beta^4 n^7}\right)e^{-\frac{6\pi R}{n\beta}} \notag \\
&~&\left. +\left(\frac{3}{2}+\frac{2\pi^2 l^2 (n^2-1)}{\beta^2 n^3}
-\frac{l^4 \pi ^4 (5 n^6-120 n^5-31 n^4+240 n^3+9635 n^2-72840 n+135111)}{90 \beta^4 n^7}\right)e^{-\frac{8\pi R}{n\beta}} \right\}
\notag \\
&~&+\frac{1}{c}\frac{2}{n-1}\left(
\frac{\pi^4 l^4 (2n^6-63n^4-672n^2+733)}{945\beta^4cn^7}e^{-\frac{4\pi R}{n\beta}}
+\frac{\pi^4 (2n^6-217n^4+1148n^2-933)}{630 \beta^4 c n^7} e^{-\frac{6\pi R}{n\beta}} \right.\notag \\
&~&\left.+\frac{l^4 \pi ^4 (2 n^6-427 n^4+13328 n^2-40320 n+27417)}{315 \beta^4 n^7}
e^{-\frac{8\pi R}{n\beta}}\right)
+O(l^4)+O(e^{-\frac{10\pi R}{n\beta}})
\eea
for $n>4$.

\subsection{Classical limit of the conformal blocks}

In the previous subsection, we claimed  that for each $f_j$, when $\mid t, j\rangle$ is a primary state in $n$-copied theory, it has no linear $c$ contribution. In this subsection, we clarify this fact from the point of view of a  large $c$ conformal block, and furthermore we show that  we only need to consider the twist vacuum module to find the linear $c$ order entanglement and R\'enyi entropies.

%As we have seen,  even though there is only one primary state in the vacuum module of normal CFT, there are more than one primary states including the twist vacuum and other primary states in the twist sector of the orbifold CFT.
Let us study the four point function between two twist operators and two vertex operators corresponding to $\mid t,j\rangle$ in the twist sector, which is primary under the conformal symmetry of the $n$-copied theory. By the conformal transformation
\be v=e^{\frac{2\pi}{\beta}(u-\frac{l}{2})}, \ee
the four-point correlation function can be transformed into
\bea \lefteqn{\langle t,j\mid {\cal{T}}^{-}(-\frac{l}{2}){\cal{T}}^{+}(\frac{l}{2}) \mid t,j \rangle \mid_u }\notag \\
&=&(\frac{2\pi}{\beta}e^{-\frac{2\pi l}{\beta}})^{\frac{c}{6}(n-\frac{1}{n})}
\langle t,j\mid {\cal{T}}^-(e^{-\frac{2\pi l}{\beta}}){\cal{T}}^{+}(1) \mid t,j \rangle \mid_v \notag,
\eea
%&=&(\frac{2\pi}{\beta}e^{-\frac{2\pi l}{\beta}})^{\frac{c}{6}(n-\frac{1}{n})}
%\sum_{p}a_p{\cal{F}}(nc,h_{\cal{T}},h_{\cal{T}},h_{t,i},h_{t,i};h_p,1-e^{-\frac{2\pi l}{\beta}}),
%\eea
%which is a sum of four point function on a $n$-copied of CFT with central charge $nc$.
which can be expanded by the conformal blocks as
\be\label{conformalblock} \sum_{p}C_{h_{\cal{T}},h_{\cal{T}}}^{h_p}C_{h_{t,j},h_{t,j}}^{h_p}
{\cal{F}}(nc,h_{\cal{T}},h_{\cal{T}},h_{t,j},h_{t,j};h_p;1-e^{-\frac{2\pi l}{\beta}}), \ee
where $nc$ is the central charge of the CFT$_n$, and $C_{h_1,h_2}^{h_3}$ is the OPE coefficient from two primary operators with $h_1$ $h_2$ conformal dimension to a primary operator with $h_3$ conformal dimension. The first four conformal dimensions in $\cal F$ are for the four external operators, two twist operators and two operators in the twisted vacuum, and the last one $h_p$ is the conformal dimension of the primary field in the propagator.
In each replica, we consider only the vacuum module in CFT$_1$, so the states in the propagator are  in the tensor product of $n$ vacuum modules, which is the trivial sector in the CFT$_n$.

One essential point is that the OPE coefficient  is of order $c^0$. The primary operators  can be normalized as
\be \langle O(z_1)O(z_2) \rangle=\frac{1}{(z_1-z_2)^{2h}}. \ee
In our case, each operator in the propagator is a combination of the stress tensors and their  partial derivatives. If  the largest number of the stress tensors  in the combination is $r$, such an operator should be normalized by  a factor of order $c^{-\frac{r}{2}}$ in the large $c$ limit. The OPE coefficient $C_{{\cal{T}},{\cal{T}}}^{h_{p}}$ equals  the expectation value in the $n$-sheeted surface. To compute it we need to transform into a full complex plane  $w$. We can decompose the transformation into two steps: the first one transforms $z$ coordinate into an $n$-sheeted fan with boundary condition
\be T^{(j)}(\tilde{w}e^{\frac{2\pi i}{n}})=T^{(j+1)}({\tilde{w}}); \ee
and the second step unfolds the $n$-sheeted fan into the full complex plane. %Because the operator we calculate is a primary operator it transform homogenously in the first step. In the second step we just need to transform
%\be T^{(j)}(\tilde{w})\rightarrow T(we^{\frac{2\pi i(j-1)}{n}})e^{\frac{4\pi i(j-1)}{n}}. \ee
In the transformations, the number of the stress tensors in the operators does not change, so the expectation value is at most order $c^0$ in the large $c$ limit, which means the OPE coefficient $C_{{\cal{T}},{\cal{T}}}^{h_{p}}$ is at most order $c^0$\cite{Chen:2013kpa}. For $C_{{t,h},{t,h}}^{h_{p}}$, we just need to insert two extra operators, and it is still of order $c^0$.

Furthermore,  the leading contribution in the conformal block  $\cal F$  is the same in the large $c$ limit. As suggested in \cite{Zamolodchikov:1985ie,Hartman:2013mia},
\be\label{conformalblock} {\cal{F}}(c,h_i,h_p,x)=\exp[-\frac{c}{6}f(h_i^{(0)},h_p^{(0)},x)+O(c^0)], \ee
where
\bea h_i^{(0)}&=&\lim_{c\rightarrow \inf}\frac{h_i}{c} \notag \\
h_p^{(0)}&=&\lim_{c\rightarrow \inf}\frac{h_p}{c}. \eea
In the case at hand, we have the relation
\be\label{pro} h_p=O(c^0) ~~~h_{t,i}-h_{t}=O(c^0), \ee
so that the classical conformal blocks are the same for all different terms in the expansion. Taking into (\ref{conformalblock}), it is easy to prove that the four-point functions are independent of $j$ in the leading order, if $\mid t,j\rangle$ is a primary operator.
Moreover, even for the two twist operators' correlation on torus, one can also only consider the twist module generated from $\mid t,j\rangle$ by Vir$^{(t)}$ for the linear $c$ order. Other modules gives only $\frac{1}{c}$ corrections. %In other words, in order to compute the leading  order contribution, we can focus on the twist vacuum module.

\subsection{Leading contribution from the twist vacuum module}

As we showed above, it is only necessary to insert the twist vacuum module to compute the leading $c$ order R\'enyi entropy.
In this subsection, we  use the Ward identity to calculate the contribution from these terms explicitly. By using the Ward identity, all of the multicorrelation functions for the descendant  operators can be derived from that for the primary operators.
From the recursion relation
\bea \lefteqn{ \langle O \mid \prod_{s=1}^m T(w_s) \prod_{t=1}^n \phi^{(t)}(z_t) \rangle} \notag \\
&=&\sum_{k=2}^m\langle O\mid (\frac{c}{2(w_1-w_k)^4}+\frac{2T(w_k)}{(w_1-w_k)^2}+\frac{\partial T(w_k)}{w_1-w_k})
\prod_{s=2\sim m,s\neq k} T(w_j) \prod_{t=1}^n \phi^{(j)}(z_j) \rangle \notag \\
&~&+\sum_{k=1}^n\langle O\mid \prod_{s=1}^m T(w_j)
(\frac{h_k\phi^{(k)}(z_k)}{(w_1-z_k)^2}+\frac{\partial\phi^{(k)}(z_k)}{w_1-z_k})
\prod_{t=1\sim n, t\neq k}\phi^{(j)}(z_j)
\rangle,
\eea
%we can derive multi-correlation function inserting any stress tensor from multi-primary operators correlation function,
where  $O$ in the bra denotes a primary operator in the infinity.
%we can insert any stress tensor in a multi-correlation function with primary operator, and it is related to the primary operators multi-correlation function by some differential on it.
With proper contour integral and contraction, we can derive any correlation function of the descendants of the primary operators.

  What we need to compute in the partition function are the ratios
\bea\label{Vir} \frac{\langle t,i^{'}\mid {\cal{T}}^-(u_1){\cal{T}}^+(u_2)\mid t,i^{'}\rangle}
{\langle t\mid {\cal{T}}^-(u_1){\cal{T}}^+(u_2)\mid t\rangle}
=\frac{\langle t,i^{'}\mid {\cal{T}}^-(z_1){\cal{T}}^+(z_2)\mid t,i^{'}\rangle}
{\langle t\mid {\cal{T}}^-(z_1){\cal{T}}^+(z_2)\mid t\rangle}
\eea
Here $\mid t,i' \rangle$ means the Virasoro descendants in the twist vacuum module generated by acting $L^{(0)}_{-m},m>0$ on the twist vacuum.
And we have changed the coordinate into a full complex plane by
\be z=e^{\frac{2\pi}{\beta}u}, \ee
so that the two inserting operators are at the origin and the infinity respectively.
By the conformal transformation and the Ward identity, all of the terms in (\ref{Vir}) can be calculated by the four-point functions, which are related to the double interval mutual information. Actually, for the simplest case, the contribution of the twist vacuum  is encoded in the correlation function of four twist operators with two of them being inserted at the origin and the infinity in the complex $z$ plane. In general, the  correlation functions of four twist operators read
\bea \lefteqn{\log <{\cal{T}}^+(z_1){\cal{T}}^-(z_2){\cal{T}}^+(z_3){\cal{T}}^-(z_4)> } \notag \\
&=&(n-1)I^{(n)}+\log<{\cal{T}}^+(z_1){\cal{T}}^-(z_4)>+\log<{\cal{T}}^-(z_2){\cal{T}}^+(z_2)> \notag \\
&=&(n-1)I^{(n)}(x,\bar{x})-\frac{c}{12}(n-\frac{1}{n})\log((z_1-z_4)(\bar{z}_1-\bar{z}_4)(z_2-z_3)(\bar{z}_2-\bar{z}_3)),
\eea
where
\be x=\frac{z_3-z_2}{z_3-z_1}\frac{z_4-z_1}{z_4-z_2} \ee
and $I^{(n)}$ is the mutual R\'enyi information.
If we set one point to infinity, then we have
\bea \lefteqn{\log <t \mid {\cal{T}}^+(z_1){\cal{T}}^-(z_2){\cal{T}}^+(z_3)>} \notag \\
&=&\log \lim_{z_4\rightarrow \inf}
z_4^{\frac{c}{12}(n-\frac{1}{n})} \bar{z}_4^{\frac{c}{12}(n-\frac{1}{n})}
 <{\cal{T}}^+(z_1){\cal{T}}^-(z_2){\cal{T}}^+(z_3){\cal{T}}^-(z_4)> \notag \\
&=&(n-1)I^{(n)}(x,\bar{x})-\frac{c}{12}(n-\frac{1}{n})\log((z_2-z_3)(\bar{z}_2-\bar{z}_3)),
  \eea
where
\be x=\frac{z_3-z_2}{z_3-z_1}. \ee
The perturbative computation of $I^{(n)}$ has been done in \cite{Chen:2013kpa,Chen:2013dxa}. We list them in Appendix D. With these results, we can derive any $f_{i^{'}}$ for the descendants of the twist vacuum.

%Setting $z_1$ to zero, we got the two twist operator correlation function in twist vacuum
%\be <t\mid {\cal{T}}^(+)(z_2){\cal{T}}^(-)(z_3) \mid t>
%=\log <t \mid {\cal{T}}^+(z_1){\cal{T}}^-(z_2){\cal{T}}^+(z_3)> \mid_{z_1=0} \ee

%With this property, we try to calculate these states contribution in partition function (We note that the definition here is different from previous section)

The first few lowest  descendants in the twist vacuum module are
\bea \mid t,1^{'}\rangle&=&(\frac{c}{12}(n-\frac{1}{n}))^{-\frac{1}{2}}L^{(0)}_{-1}\mid t\rangle \notag \\
\mid t,2,1^{'}\rangle &=&(c(\frac{2}{3}n-\frac{1}{6n}))^{-\frac{1}{2}}L^{(0)}_{-2}\mid t\rangle \notag \\
\mid t,2,2^{'}\rangle &=&m(L^{(0)}_{-1}L^{(0)}_{-1}+aL^{(0)}_{-2})\mid t\rangle \eea
where
\bea a&=&-\frac{3(n^2-1)}{8n^2-2} ,\notag \\
 m&=&\left(\frac{c^2(n^2-1)^2}{72n^2}+c\frac{(n^2-1)(5+7n^2)}{24n(4n^2-1)}\right)^{-\frac{1}{2}}. \eea
Note that these states consist of a special set of excited states in the twist sector, which are the decendants of twist vacuum generated by $L^{(0)}_{-m}$.
And their contributions to the leading linear $c$ order are, respectively,
\bea \tilde{f}_{1^{'}}&=&\frac{\langle t,1'\mid {\cal{T}}^{-}(-\frac{l}{2}){\cal{T}}^{+}(\frac{l}{2})\mid t,1 '\rangle}
{\langle t\mid {\cal{T}}^{-}(-\frac{l}{2}){\cal{T}}^{+}(\frac{l}{2})\mid t \rangle} \notag \\
&=&c(n^2-1)^3\left(\frac{\pi^4}{108\beta^4n^5}l^4
+\frac{(-11-2n^2+19n^4)\pi^6}{9720\beta^6n^9}l^6+O(l^7)\right)+(1+\frac{(n^2-1)\pi^2}{3\beta^2n^2}l^2+O(l^4))
+O(c^{-1}) \notag \\
 \tilde{f}_{2,1^{'}}&=&\frac{\langle t,2,1'\mid {\cal{T}}^{-}(-\frac{l}{2}){\cal{T}}^{+}(\frac{l}{2})\mid t,2 ,1'\rangle}
{\langle t\mid {\cal{T}}^{-}(-\frac{l}{2}){\cal{T}}^{+}(\frac{l}{2})\mid t \rangle} \notag \\
&=&c(-1 + n^2)^2\left(\frac{(-1+4n^2)\pi^4}{54\beta^4n^5}l^4
+\frac{(11-9n^2-111n^4+289n^6)\pi^6}{4860\beta^6n^9}l^6+O(l^7)\right)+O(c^0) \notag \\
 \tilde{f}_{2,2^{'}}&=&\frac{\langle t,2,2'\mid {\cal{T}}^{-}(-\frac{l}{2}){\cal{T}}^{+}(\frac{l}{2})\mid t,2,2' \rangle}
{\langle t\mid {\cal{T}}^{-}(-\frac{l}{2}){\cal{T}}^{+}(\frac{l}{2})\mid t \rangle} \notag \\
&=&c^2 (\frac{(n^2-1)^6\pi^8}{23328\beta^8n^{10}}l^8+O(l^9))+
c(-1+n^2)^3\left(\frac{\pi^4}{54\beta^4n^5}l^4
+\frac{(-11-32n^2+49n^4)\pi^6}{4860\beta^6n^9}l^6+O(l^7)\right)+O(c^0). \notag \\ \eea
%Up to order $l^6$,  the result is in perfect match with previous calculation.
The leading $c$ contribution for the R\'enyi entropy reads
\bea S_n^{\mbox{(class)}}&=&\frac{\pi cR}{6\beta}\frac{n+1}{n}
+c\frac{n+1}{6n} \log l +c(n+1)\left(\frac{\pi^2}{36n^3\beta^2}l^2-\frac{\pi^4(n^6+9n^4-9n^2+11)}{12960n^7\beta^4}l^4 \right. \notag \\
&~&\left.+\frac{\pi^6(19n^{10}+85n^8-125n^6+251n^4-274n^2+188)}{2449440n^{11}\beta^6}l^6\right) \notag \\
&~&+c(-1+n)^2(1+n)^3\left(-\frac{\pi^4}{54\beta^4n^5}l^4
-\frac{(-11-2n^2+19n^4)\pi^6}{4860\beta^6n^9}l^6\right)e^{-\frac{2\pi R}{\beta}} \notag \\
&~&+c(-1+n)(1+n)^2\left(-\frac{(3n^2-1)\pi^4}{18\beta^4n^5}l^4
-\frac{(11+n^2-101n^4+209n^6)\pi^6}{1620\beta^6n^9}l^6\right)e^{-\frac{4\pi R}{\beta}} \notag \\
&~&+O(l^7)+O(e^{-\frac{6\pi R}{\beta}}), \notag \\ \eea
and
\be\label{SEE1} S_{EE}^{\mbox{(class)}}=\lim_{n\rightarrow 1}S_n^{\mbox{class}}=
\frac{\pi cR}{3\beta}+c\left(\frac{\pi^2 l^2}{18\beta^2}-\frac{\pi^4 l^4}{540\beta^4}+\frac{\pi^4 l^6}{8505\beta^4}\right)+O(l^8), \ee
for leading order of the expansion.

From the result, we find that in the entanglement entropy there is no finite size correction proportional to the powers of  $e^{-\frac{2\pi R}{\beta}}$. Such a correction, if it existed,  should come from the 4-point functions of the descendants
\be \langle t,i,\bar{i} \mid {\cal{T}}^{-}(u_1) {\cal{T}}^{+}(u_2) \mid t,i ,\bar{i} \rangle \mid_u
=\langle t,i,\bar{i} \mid {\cal{T}}^{-}(z_1) {\cal{T}}^{+}(z_2) \mid t,i, \bar{i} \rangle \mid_z, \ee
where $\mid t,i,\bar{i} \rangle$ is a  state generated by a set of creation generators $L^{(0)}_{-r}$ $\tilde{L}^{(0)}_{-r}$ acting on the twist vacuum $\mid t\rangle$. Consider the Ward identity
\be\label{ward1} [L_r,{\cal{T}}^{-}(u_1,\bar{u}_1)]=(r+1)h_{\cal{T}}u_1^r{\cal{T}}^{-}(u_1,\bar{u}_1)+
u_1^{r+1}\partial {\cal{T}}^{-}(u_1,\bar{u}_1), \ee
where
\be h_{\cal{T}}=\frac{c}{24}(n-\frac{1}{n}). \ee
In the right side of (\ref{ward1}),  because $h_{\cal{T}}=\frac{c}{24}(n-\frac{1}{n})$, the first term is of order $n-1$; because the four point function is constant when $n=1$, the second term should also be of order $n-1$. Similarly when moving $L_{-r}$ to the left side, the commutation term will contribute an $n-1$.
To calculate the correlation function, we move all of the annihilation operators to the right side and then move the reduced creation operators to the left side. %When moving the annihilation operators to the right side there are a lot of contraction as the right hand side of the equation \ref{wald1}, which is of order $n-1$. For a term with at least one contraction from moving annihilation operators to right side, which contribute at least a $n-1$ factor, the creation operators can not be annihilate by the annihilation operators moved to left side, so we need to move the reduced creation operators to left side which contribute at least another $n-1$ factor. For each term with contraction when moving annihilation operator to the right side, it is at least of order $(n-1)^2$. There is only one term which has no contraction, in which we move all of the annihilation operators to the right side. Considering the normalization, this term is of order $c^0$. So for R\'enyi entropy in linear c order the contribution from descendants is
It turns out that the leading contribution terms are at least of order $(n-1)^2$. Therefore
\bea S_n^{(\mbox{class})}&=&-\frac{1}{n-1}\left\{\frac{\pi Rc}{6n\beta}-\frac{\pi Rcn}{6\beta}+
\log \langle t\mid {\cal{T}}^{-}(u_1){\cal{T}}^{+}(u_2)\mid t\rangle \mid_u\right\}+O(n-1) \notag \\
&=&-\frac{1}{n-1}\left\{\frac{\pi Rc}{6\beta}(\frac{1}{n}-n)+
\log\left( (\frac{2\pi}{\beta})^{\frac{c}{6}(n-\frac{1}{n})}
\langle t\mid {\cal{T}}^{-}(z_1){\cal{T}}^{+}(z_2)\mid t\rangle \mid_z\right)\right\}+O(n-1) \notag \\
&=&\frac{(n+1)\pi Rc}{6n\beta}-\frac{1}{n-1}
\left\{\log \left((\frac{2\pi}{\beta})^{\frac{c}{6}(n-\frac{1}{n})}\frac{1}{\mid z_1-z_2 \mid^{\frac{c}{6}(n-\frac{1}{n})}}\right) +O(n-1)^2\right\}+O(n-1) \notag \\
&=&\frac{(n+1)\pi Rc}{6n\beta}+\frac{(n+1)c}{6n}\log \sinh \frac{\pi l}{\beta} +O(n-1). \label{linearc}
\eea
In the third equation we use the classical conformal block \cite{Hartman:2013mia}, and the entanglement entropy is
\be S_{EE}^{(\mbox{class})}=\frac{\pi cR}{3\beta}+\frac{c}{3} \log \sinh \frac{\pi l}{\beta}, \ee
which matches with previous result (\ref{SEE1}) up to order $l^6$.
This is the high temperature entanglement entropy for a large interval and it satisfies the relation
\be \lim_{\epsilon \rightarrow 0} S_{EE}(R-\epsilon)-S_{EE}(\epsilon)=S_{th}. \ee

\section{Holographic R\'enyi entropy}

In this section, let us calculate the entanglement entropy and the R\'enyi entropy holographically up to  1-loop order. In the field theory, by the replica trick the R\'enyi entropy can be transformed into the partition function on a higher genus Riemann surface. Holographically, this partition function can be computed in the semi-classical AdS gravity in the large central charge limit. Based on the AdS/CFT correspondence the gravity configurations must be  the classical solutions with the asymptotically boundary being the   Riemann surface \cite{Krasnov:2000zq}. Moreover,
for the same Riemann surface, there may be more than one gravitational solution. The partition function is the summation of the classical contributions and the quantum corrections at different saddle points.
Among different saddle points, the one with the smallest action dominates the contribution, and other saddle points give non-perturbative corrections of order $e^{-c}$. Therefore, in the large $c$ limit we only need to consider the saddle point with the smallest action. The regulated on-shell action of this saddle point gives the classical contribution, corresponding to the leading linear $c$ result in the field theory, while the 1-loop determinant of the fluctuations around the saddle point gives the  quantum correction, which corresponds to the order $c^0$ results in the field theory.

As in \cite{Faulkner:2013yia}, we assume that only the handle-body solutions contribute to the partition function. The handle-body solutions could be obtained by extending the Schottky uniformization of the Riemann surface to the bulk.
In this section we first give a brief review on the Schottky uniformization and the on-shell action. Then we  discuss the monodromy condition for the $n$-sheeted torus pasted along a single large interval to find the uniformization. We compute the classical part of the holographic R\'enyi entropy perturbatively. Furthermore, after carefully studying the primitive class of the Schottky group, we calculate the 1-loop corrections to the entropies, following the treatment in  \cite{Yin:2007gv}.

\subsection{Schottky uniformization and the partition function}

In three dimensional AdS pure gravity, all solutions with constant negative curvature are quotients of the AdS space.  In terms of  the Poincar\'e coordinates,   the AdS space could be described as an upper-half space with the metric,
\be ds^2=\frac{dud\bar{u}+d\xi^2}{\xi^2}, \hs{3ex}\xi \geq 0,\ee
where $u=x+iy$ is the coordinate of a complex plane.
The isometry group of AdS$_3$ is $PSL(2,C)$ \cite{Maloney:2007ud}. The coordinates $u,\xi$ can be combined into  a quaternion $\zeta=u+j\xi$, on which  the isometry group acting as
\be \tilde{\zeta}=\frac{a\zeta+b}{c\zeta+d}, \ee
with $ad-bc=1, a, b, c, d \in C$.  At the asymptotic boundary $\xi\rightarrow 0$, the transformation is just a linear M\"obius transformation on a complex plane,
\be u\rightarrow \frac{au+b}{cu+b}. \ee
Generally, the gravity solution can be written as $AdS_3^{'}/\Gamma$, where $AdS_3^{'}$ is the global $AdS_3$ with some fixed points being removed and $\Gamma$ is the discrete subgroup of $PSL(2,C)$. The asymptotic boundary is $\Omega/\Gamma$, where $\Omega$ is a full complex plane with some fixed points being removed.
If we focus on the handle-body solutions, the subgroup is just the Schottky group.
 %Then we get a constant negative curvature gravity solution with $\Omega/\Gamma$ as asymptotic boundary condition. Furthermore $\Omega/\Gamma$ can be regarded as one uniformization of the Riemann surface. To find the bulk solution, we only need to find the Uniformization of the boundary Riemann surfacea and extend the uniformization to the bulk.

Every compact Riemann surface can be realized by the Schottky uniformization. For a genus-$g$ Riemann surface $X$, its fundamental group $\pi_1(X)$ is generated by $2g$ generators,
\be a_1,~a_2,~...~a_g;~b_1,~b_2,~...~b_g \ee
with constraints
\be \prod_{i=1}^{g}[a_i,b_i]=1,~~~[a,b]=aba^{-1}b^{-1}. \ee
One can always choose $g$ loxodromic generators $L_1,~L_2,~...~L_g$ and a  fundamental regain $D$ bounded by $2g$ circles $C_1,~C_2,~...~C_g$ and $C^{'}_1,~C^{'}_2,~...~C^{'}_g$, such that $L_i(C_i)=C^{'}_i$. Identifying $g$ pairs of circles by the generators, we obtain a quotient space, which is just a genus-$g$ Riemann surface. Here $a_i$ is just the image of $C_i$($C_i^{'}$) under the quotient map in the homology group; the group of covering $\Omega \rightarrow \Omega/\Gamma$ is the smallest normal subgroup ${\cal{N}}$ containing the elements $a_i$'s, and the Schottky group is isomorphic to $\pi_1(X)/{\cal{N}}$. The Schottky uniformization can be extended to the bulk, which is an automorphism of the $AdS_3$ space, with  the $a_i$ cycles in the bulk being contractable.

For one Riemann surface, there are more than one way to choose the $a_i$ and $b_i$ cycles. Different choices of the generators of the fundamental group correspond to different realizations of  the Schottky uniformization. Even though different Schottky uniformizations describe the same Riemann surface, their extensions to the bulk give different gravity solutions.

The Schottky uniformization problem for a general Riemann surface could be solved by  considering the differential equation
\be \label{pde}
\psi''(u)+\frac{1}{2}R^{(S)}(u)\psi(u)=0
\ee
where $R^{(S)}(u)$ is the Schottky projective connection on a marked Riemann surface. $R^{(S)}(u)$ is uniquely determined by the normal subgroup ${\cal{N}}$. Namely it depends on the choice of the generators. A ratio of the linearly independent solutions of the above equation determines the quotient map in the covering space $\Omega$. More importantly,  it turns out that up to a normalization $R^{(S)}$ is just the holomorphic stress tensor $T(u)$ of the Liouville theory \cite{Krasnov:2000zq},
 which is the regulated on-shell action of the bulk solution of the AdS$_3$ gravity. %Actually, with proper normalization, $T(u)=\frac{c}{12}R^{(S)}(u)$.
 The explicit forms of the stress tensor depend on $(3g-3)$ complex accessory parameters with respect to the holomorphic quadratic differentials on the Riemann surface such that the determination of the uniformization map is usually a very difficult problem.  However, for the Riemann surface in computing the R\'enyi entropy, the uniformization problem could be solved perturbatively in some cases due to the replica symmetry. For the double interval case\cite{Faulkner:2013yia} , the stress tensor takes the form
\be R^{(S)}(u)=\sum_i\frac{\Delta}{(u-u_i)^2}+\frac{\gamma_i}{u-u_i}, \ee
where
\be \Delta=\frac{1}{2}(1-\frac{1}{n^2}), \ee
 and there is only one conformal invariant accessory parameter. For the single interval on a torus, the stress tensor takes the form \cite{Barrella:2013wja}.
 \be R^{(S)}(u)=\sum_i(\Delta \wp(u-u_i)+\gamma_i \zeta(u-u_i))+\delta, \ee
 where $\wp $ is the doubly periodic  Weierstrass function
\bea \wp(u;R,\frac{i}{T})&=&\frac{1}{u^2}+\sum_{(m.n) \neq (0,0)}
[\frac{1}{(u+mR+\frac{in}{T})^2}-\frac{1}{(mR+\frac{in}{T})^2}] \\
&=&\sum_{m}\frac{\pi^2T^2}{\sinh^2[\pi T(u+mR)]}
-\sum_{m \neq 0}\frac{\pi^2T^2}{\sinh^2(\pi mTR)}+\frac{\pi^2T^2}{3}, \eea
and
\be \zeta(u)=\sum_m \pi T \coth[\pi T(u+mR)]
+\sum_{m\neq 0}\frac{\pi^2T^2u}{\sinh^2\pi mTR}-\frac{\pi^2 T^2u}{3}. \ee
To solve the problem, one has to impose the monodromy condition on some cycles to fix the accessory parameters. The different choices on the cycles with trivial monodromy give different Schottky uniformization.

On the other hand, the regulated on-shell action of a AdS$_3$ gravity solution is the so-called  Takhtajan-Zograf action\cite{Krasnov:2000zq}. Moreover, the dependence of the action on the moduli parameter has been studied in \cite{Zagraf:1988, Takhtajan:2002cc}. For the gravitational configuration dual to the $n$-sheeted Riemann surface,  the action obeys the equation \cite{Faulkner:2013yia}
\bea
 \frac{\partial S_n}{\partial u_i}&=&-\frac{cn}{6(n-1)}\gamma_i. \label{partialm}
\eea
This equation allows us to obtain the classical action of the gravity solution corresponding to a Schottky uniformization. Among  different uniformizations for the same Riemann surface, the one leading to the least gravitational action dominates the partition function.

%Each of the them are differential equations on higher genus Riemann surface, and the solution is not necessary single valued. We can tune the parameter, such that there are g (g is the genus number for the Riemann surface) trivial cycles (for example $a_i$ cycles), along which the solution is single valued.
%Choosing two independent solutions $\psi_1$ and $\psi_2$, the fraction
%\be \xi=\frac{\psi_1(u)}{\psi_2(u)}, \ee
%is the coordinate in $\Omega$ to realize the Schottky Uniformization. As we see, along the $a_i$ cycles $w$ don't change, while along $b_i$ cycles, w has an $SL(2,C)$ transformation, which is Schottky Uniformization.

Here let us  focus on the case that there is a single interval on a torus. Because the cycles around two branch points are always of trivial monodromy, we have
\be \gamma_1=-\gamma_2. \ee
For convenience, we redefine the functions and rewrite the stress tensor as
\be\label{mT}
R^{(S)}(u)=\Delta\tilde{\wp}(u-u_1)+\Delta\tilde{\wp}(u-u_2)+\gamma\tilde{\zeta}(u-u_1)-\gamma\tilde{\zeta}(u-u_2)
+\tilde{\delta}, \ee
where
\bea \tilde{\wp}(u)&=&\sum_{m=-\inf}^{\inf}\frac{\pi^2T^2}{\sinh^2[\pi  T(u+mR)]}, \notag \\
 \tilde{\zeta}(u)&=&\sum_{m=-\inf}^{\inf}\pi T\coth[\pi T(u+mR)].\eea
For the classical partition function, we need to calculate the on-shell action of the gravity solution with proper  boundary terms as regulators. It turns out Eq. (\ref{partialm}) is not enough to determine the action completely.  Besides the dependence of the action on the accessory parameter, we have to take into account  its dependence on the size of the torus. In \cite{Chen:2014unl}, we proposed  another differential relation on the partition function, in addition to (\ref{partialm}),  in order to determine the  size dependence of the partition function completely
\bea \label{partialc}
 \frac{\partial S_n}{\partial R}&=&\frac{c}{12\pi}\frac{n}{n-1}\beta(\tilde{\delta}-\tilde{\delta}_{n=1}). \eea
  With Eqs. (\ref{partialm}) and (\ref{partialc}), we can determine the partition function completely.

%As we said previously, there can be different monodromy choice which correspond to different classical solution, and in large c limit, the one with smallest action dominate the contribution, while others' have non-perturbative correction of order $O(e^{-\frac{1}{c}})$. For the torus, small interval and large interval has different dominate solutions. We first review the monodromy condition for small interval \cite{Barrella:2013wja}. Then we introduce our new monodromy condition, which has smaller action for large interval.

From the holographic entanglement entropy of one single interval in the black hole background \cite{Azeyanagi:2007bj}, there should be a phase transition when the interval becomes  large enough. This means that for a very large interval one should impose a different set of monodromy conditions, which leads to different Schottky uniformization. To support our choice on the monodromy conditions for the large interval case, we will compute  in the following the holographic entanglement entropy and compare it with the result in \cite{Azeyanagi:2007bj}.
%For the small interval, the $n$-sheeted horizonal cycles and upright cycles are a set of canonical bases. We must choose the $n$ upright cycles along the thermal direction as trivial cycles. For the large interval, we should choose another set of  trivial cycles: one cycle goes across $n$-sheeted torus passing across the branch cut and the other $n-1$ independent cycles goes around the complementary part of the trivial cycles. To show the difference of two choices, we first solve the monodromy condition for $n \rightarrow 1$ and calculate the entanglement entropies for different choices.

Let us first review the holographic computation in the short interval case. We set the branch cut at $[u_1,u_2]$. At a high temperature,  the thermal cycle should be  of trivial monodromy, and the wave function transforms as
\be\label{monodromy1} \psi(u+\frac{i}{T})=-\psi(u). \ee
If we transform into the $z$ coordinate, there is no minus sign in the monodromy condition. As discussed in \cite{Barrella:2013wja}, to compute the holographic entanglement entropy we only need to study the solution near $n=1$ and expand the wave function and the parameters with respect to $n-1$ as
\be \psi(u)=\psi^{(0)}(u)+\sum_{i=1}^{\inf}(n-1)^i\psi^{(i)}(u), \ee
\be \gamma=\sum_{i=1}^{\inf}(n-1)^i\gamma^{(i)}, \ee
\be \tilde{\delta}=-2\pi^2T^2+\sum_{i=1}^{\inf}(n-1)^i\tilde{\delta}^{(i)}, \ee
with
\be\label{zero1} \psi^{(0)}(u)=Ae^{\pi Tu}+Be^{-\pi Tu}. \ee
Expanding the trivial monodromy condition with respect to $(n-1)$, we have
\be \psi^{(i)}(u+\frac{i}{T})=-\psi^{(i)}(u), \ee
at each order.
With proper redefinition of $A$ and $B$ in Eq. (\ref{zero1}), we can also set
\be\label{mon} \psi^{(j)}(0)=\psi'^{(j)}(0)=\psi^{(j)}(\frac{i}{T})=\psi'^{(j)}(\frac{i}{T})=0,~\mbox{for}~j=1,2,... \ee
Taking the expansions of the wave function and the parameters into the equation, we find the following equation at the leading order
\be\label{leadingorder} \psi^{(1)''}(u)-\pi^2T^2\psi^{(1)}(u)+\frac{1}{2}m(u)\psi^{(0)}(u)=0, \ee
where
\be m(u)=\tilde{\wp}(u-u_1)+\tilde{\wp}(u-u_2)
+\gamma^{(1)} \tilde{\zeta}(u-u_1)-\gamma^{(1)} \tilde{\zeta}(u-u_2)+\tilde{\delta}^{(1)}.\ee
We get the solution
\be \psi^{(1)}(u)=\frac{e^{-u\pi T}}{2\pi T}\int_0^u e^{r\pi T}m(r)\psi^{(0)}(r)dr
-\frac{e^{u\pi T}}{2\pi T}\int_0^{u}e^{-r\pi T}m(r)\psi^{(0)}(r)dr. \ee
Furthermore, considering the last two equations in (\ref{mon}), we get
\bea \int_0^1m(\frac{is}{T})ds=0,
 \hs{3ex}\int_0^1e^{2\pi is}m(\frac{is}{T})ds=0. \eea
Solving these equations, we find
\bea \tilde{\delta}^{(1)}&=&0, \notag \\
 \gamma^{(1)}&=&2\pi T \coth \pi T(u_2-u_1). \eea
Taking  $\tilde{\delta}$ and $\gamma$ into Eqs. (\ref{partialm},\ref{partialc}), we obtain the classical entanglement entropy
\be S_{EE}^{(\mbox{classical})}=\frac{c}{6}\log \sinh^2\pi T(u_2-u_1)+\mbox{const}, \ee
which is the geodesic length in the bulk connecting two branch points.

On the other hand, for the large interval case, we choose another $n$ cycles of trivial monodromy. To compare with the small interval case, we set the branch cut at $[0,u_1]\bigcup[u_2,R]$.  Among $n$ trivial cycles, there is one cycle that goes across the branch cut for $n$ times. This cycle  is denoted as $\tilde{A}^{(1)}$ in Fig. 2 and  Figs. \ref{cycle2}. There are other $n-1$ independent cycles enclosing $[u_1,u_2]$, the complementary part of the large interval. These cycles are denoted as   $\tilde{A}^{(i)},i>1$ in Fig. \ref{cycle2}.   As in the small interval case, we expand the wave function and parameter with respect to $(n-1)$,
\be \psi=\psi^{(0)}(u)+\sum_{i=1}^{\inf}(n-1)^{i}\psi^{(i)}(u), \ee
\be \tilde{\delta}=-\frac{2\pi^2T^2}{n^2}+\sum_{i=1}^{\inf}(n-1)^{i}\tilde{\delta}^{(i)}, \ee
\be \gamma=\sum_{i=1}^{\inf}(n-1)^i \gamma^{(i)}, \ee
%\be \psi_{r}(z)=\sum_{i}(n-1)^i\psi^{(i)} \ee
and the zeroth order wave function is
\be \psi^{(0)}=A^{'}e^{\frac{\pi T}{n}u}+B^{'}e^{-\frac{\pi T}{n}u}. \ee
The monodromy condition for the cycle $\tilde{A}^{(1)}$ is
\be \psi(u+\frac{in}{T})=-\psi(u). \ee
where $u+\frac{in}{T}$ means that the argument goes across the branch cut for $n$ times. Expanding the wave function with respect to $n-1$, we find the same differential equation (\ref{leadingorder})  at the leading order, and  the same $\tilde{\delta}^{(1)}$ and $\gamma^{(1)}$ as in the short interval case. The only difference comes from
$\tilde{\delta}^{(0)}=-\frac{2\pi^2T^2}{n^2}$.
Taking them into Eqs. (\ref{partialm}) and (\ref{partialc}), we obtain the classical entanglement entropy of the large interval
\be S_{EE}^{(\mbox{classical})}=\frac{c}{6}\log \sinh^2\pi T(u_2-u_1)+\frac{c}{3}\pi TR. \ee
Namely, the holographic entanglement entropy(HEE) of a very large interval is the sum of HEE of its complementary interval and the horizon length of a static BTZ black hole. This is exactly the result  suggested in \cite{Azeyanagi:2007bj}.

\subsection{Holographic R\'enyi entropy: Classical part}

In this subsection  we develop a systematic way to solve the monodromy problem and calculate holographically the classical  R\'enyi entropy for the large interval on a circle at high temperature. We need to solve the equation
\be \psi^{''}(u)+\frac{1}{2}R^{(S)}(u)\psi (u)=0, \ee
by tuning the parameters $\gamma$ and $\tilde{\delta}$ such that the two solutions for the second order differential equation have trivial monodromy  along the appointed cycles $\tilde{A}^{(i)}$. For convenience, we take a conformal transformation
\be z=e^{\frac{2\pi u}{\beta}}, \ee
and define some useful parameters
\be z_R=e^{-2\pi TR}, ~~~z_1=e^{2\pi Tu_1} ,~~~z_2=e^{2\pi Tu_2}. \ee
The torus is transformed into a solid annulus with the inside and outside circles being identified. The branch cut is at $[1,z_1]\bigcup[z_2,\frac{1}{z_R}]$, the $\tilde{A}^{(1)}$ cycle is the one that goes around the origin for $n$ times, and the $\tilde{A}^{(i)},i>1$ cycles are those enclosing $[z_1,z_2]$ in different sheets. With the conformal transformations,
\be z\rightarrow z\cdot z_R^m, \ee
we can cover the full complex plane with a series of annuli.%, with this picture we can study the conformal transformation of the wave function.

To study the monodromy problem for the large interval and high temperature case analytically, we can take a Laurent or Taylor expansion of the wave function about the origin and branch point and also take an expansion with respect to some parameters. Let us first consider the expansion about the origin. Because of the monodromy condition, we may rewrite the wave function as
\be \psi^{(1,\pm)}=z^{\frac{1}{2}(1 \pm \frac{1}{n})}f^{(\pm)}(z), \ee
%Solving the equation we can get $\delta$ and solution for f depending on $\gamma$
%We can assume
where $f^{(\pm)}$ should be single valued in the region  $z_2z_R<|z|<z_1$ with its Laurent expansion being convergent.
Assuming the wave function and the parameters can be expanded with respect to $z_R$ and $z$ as
\be\label{wave1} f^{(\pm)}(z,z_R,z_1,z_2)=\sum_{r=0}^{\inf}f_r^{(\pm)}(z,z_1,z_2)z_R^r, \ee
\be \gamma=\sum_{r=0}^{\inf}\gamma_rz_R^r, \hs{3ex} \delta=\sum_{r=0}^{\inf}\delta_rz_R^r, \ee
\be f^{(\pm)}_r(z,z_1,z_2)=\sum_{n=-r}^{\inf}f^{(\pm)}_{r,n}(z_1/z_2)(z/z_2)^n, \ee
with the normalization as
\bea &~&f^{(\pm)}_0=1 \notag \\
 &~&f^{(\pm)}_{r,0}=0~~~\mbox{for}~r>0, \eea
we find that $f^{(\pm)}_{r,n}$ and $\delta_r$ can be solved order by order depending on $\gamma_r$. As $z_1\simeq z_2$, the coefficients $f_{r,n}^{(\pm)}$ are of order 1 and
\be z_R^rz_2^{-n}z^n=(\frac{z_2z_R}{z})^r(\frac{z}{z_2})^{r+n},\hs{3ex} ~~~r\geq 0 ,~n\geq -r, \ee
so the wave function should be convergent in the region  $z_2z_R<z<z_1$.

Next we consider the wave function around $[z_1,z_2]$. The wave function around $z_1$ and $z_2$ can be written as
\be \psi^{(2,\pm)}=(z-z_1)^{\frac{1}{2}(1\pm\frac{1}{n})}(z-z_2)^{\frac{1}{2}(1\mp\frac{1}{n})}g^{(\pm)}(z,z_1,z_2,z_R) .\ee
The prefactors encode the information of the monodromy, and $g^{(\pm)}$ is single valued around each of the branch points so that it is analytic in the region
$|z-z_2|<\min\{|z_2z_R-z_1|,|\frac{z_1}{z_R}-z_2|\}$. In this region, the wave function is convergent,  and the function $g^{(\pm)}$ can be expanded
 with respect to $z_R$, $(z-z_2)$ and $(z_1-z_2)$,
\be g^{(\pm)}(z,z_1,z_2,z_R)=\sum_{r=0}^{\inf}g^{(\pm)}_r(z,z_1,z_2)z_R^r \ee
\be g^{(\pm)}_r(z,z_1,z_2)=\sum_{n=0}^{\inf}g^{(\pm)}_{r,n}(z_1,z_2)(z-z_2)^n \ee
\be g^{(\pm)}_{r,n}(z_1,z_2)=\sum_{m=0}^{\inf}g^{(\pm)}_{r,n,m}(z_2)(z_1-z_2)^m, \ee
with normalization
\bea g^{(\pm)}_{0,0}&=&1 \notag \\
g^{(\pm)}_{r,0}&=&0 ~~~\mbox{for}~r>0, \eea
and the parameters can be expanded as
\bea &~&\gamma_0=\sum_{n=-1}^{\inf}\gamma_{0,n} (z_1-z_2)^n, \notag \\
 &~& \gamma_r=\sum_{n=0}^{\inf}\gamma_{r,n} (z_1-z_2)^n, ~\mbox{for}~r>0 \notag \\
&~& \delta_r=\sum_{n=0}^{\inf}\delta_{r,n} (z_1-z_2)^n. \eea
Taking in the previous  result on $\delta_r$, we can solve all of the parameters and coefficients order by order.
We list the solutions of $\gamma$ and $\delta$ to the first few lowest orders  in Appendix E.  Integrating Eqs. (\ref{partialm}) and (\ref{partialc}) we get the classical part of $n$th R\'enyi entropy
\bea S_n&=&\frac{c}{6}\pi TR\frac{n+1}{n}+\frac{c (n+1)}{6 n} \log (u_1-u_2)+ c (n+1)(\frac{\pi ^2 T^2}{36n^3}
(u_1-u_2)^2 \notag \\
&~&-\frac{\pi ^4  \left(n^6+9 n^4-9 n^2+11\right) T^4}{12960 n^7}(u_1-u_2)^4+ \notag \\
&~&\frac{\pi ^6 \left(19 n^{10}+85 n^8-125 n^6+251 n^4-274 n^2+188\right) T^6}{2449440 n^{11}}(u_1-u_2)^6
\notag \\
&~&-\frac{\pi ^8 }{5878656000 n^{15}} (4187 n^{14}+12787 n^{12}-22521 n^{10}+55919 n^8-100079 n^6+143481 n^4
\notag \\
&~&-126067 n^2+58213) T^8 (u_1-u_2)^8) \notag \\
&~&+c (n-1)^2 (n+1)^3\left( -\frac{\pi ^4  T^4}{54 n^5}(u_1-u_2)^4
-\frac{\pi ^6 \left(19 n^4-2 n^2-11\right) T^6}{4860 n^9}(u_1-u_2)^6 \right. \notag \\
&~&\left.-\frac{\pi ^8  \left(4187 n^8-572 n^6-4218 n^4-1412 n^2+4607\right) T^8}{12247200 n^{13}}(u_1-u_2)^8\right)e^{-2\pi TR} \notag \\
&~&+c (n-1) (n+1)^2\big(-\frac{\pi ^4  \left(3 n^2-1\right) T^4}{18 n^5}(u_1-u_2)^4  \notag \\
&~&-\frac{\pi ^6 \left(209 n^6-101 n^4+n^2+11\right) T^6}{1620 n^9}(u_1-u_2)^6  \notag \\
&~&-\frac{\pi ^8 }{4082400 n^{13}} (180041 n^{10}-104551 n^8-17506 n^6+43798 n^4
\notag \\
&~&-6455 n^2-4607) T^8(u_1-u_2)^8 \big)e^{-4\pi TR}+O((u_1-u_2)^{10})+O(e^{-6\pi TR}).
\eea
Recalling $u_1-u_2=l$, we find that the classical HRE $S_n$ is in complete agreement with the field theory result (\ref{linearc}) up to order $l^6$ and $e^{-4\pi TR}$.

%The result match exactly with the CFT's computation.

\subsection{Holographic R\'enyi entropy: 1-loop correction}

In the previous subsection, we calculated the on-shell action for the gravity solution, which gives the classical part of the holographic  R\'enyi entropy and entanglement entropy. In this section, we derive the 1-loop quantum correction to the holographic R\'enyi entropy by computing the functional determinants for the fluctuations around  the corresponding classical background. As proposed in \cite{Maloney:2007ud}, for a handle-body solution  realized as quotient space by a Schottky group $\Gamma$ in pure AdS$_3$ gravity, the 1-loop partition function is given by \cite{Yin:2007gv}
\be Z_{1-loop}=-\sum_{\gamma\in {\cal{P}}}\sum_{m=2}^{\inf}\log(1-q_{\gamma}^m), \ee
where ${\cal{P}}$ denotes the primitive conjugate class of the Schottky group, and $q_{\gamma}^{\pm\frac{1}{2}}$ denotes the eigenvalues of the Schottky group element $\gamma\subset SL(2,C)$, with $|q_{\gamma}|<1$. A group element is primitive if it cannot be written as $\gamma=\beta^n$ for $n>1$.

To read the 1-loop partition function, we need to find the corresponding $SL(2,C)$ elements for every Schottky generator and the primitive elements constructed from them. To study the corresponding $SL(2,C)$ elements in the Schottky group, we need to study the monodromy around the cycles. We can solve the wave function in different charts covering the Riemann surface.  If two charts have an overlap, there is an  $SL(2,C)$ transformation between the solutions in the overlap. For each cycle there are a series of charts covering  it so that the Schottky group elements are  the multiplication of a series of $SL(2,C)$ transformations. The crucial point is to study the $SL(2,C)$ transformation between the solutions in two overlapping charts. Since we only want to calculate the lowest order terms with respect to the modular parameters, we may expand different wave functions in the overlap, such that to each order there are only finite number of terms in the expansion. Comparing the coefficients of two different expansions, we can read the $SL(2,C)$ transformation to the fixed order.

Since $\psi^{(1,\pm)}$ is expanded at $z=0$ and $\psi^{(2,\pm)}$ is expanded at $z=z_2$, there is no direct way expanding the two wave function in the same region. In the case at hand, to study the Schottky transformation we need another wave function,  connecting $\psi^{(1)}$ and $\psi^{(2)}$. For convenience, we will write the wave function in the new coordinate %by a conformal transformation
\be\label{ztot} t=t_0\frac{z+e^{\pi T(u_1+u_2)}}{z-e^{\pi T(u_1+u_2)}}, \ee
with
\be t_0=\frac{e^{\pi T(u_2-u_1)}-1}{e^{\pi T(u_2-u_1)}+1},  \ee
which set
\bea &~&z=0 \rightarrow t=-t_0, \notag \\
&~&z=\inf \rightarrow t=t_0. \eea
The new wave function can be expanded  as
\be\label{wave3} \psi^{(3)\pm}(t)=(t+t_0)^{\frac{1}{2}(1 \pm \frac{1}{n})}(t-t_0)^{\frac{1}{2}(1 \mp \frac{1}{n})}
\sum_{m=0}^{\inf}\frac{1}{(t+t_0)^m(t-t_0)^m}h_m(t,t_0)z_R^m, \ee
with
\bea h_0(t,t_0)&=&\sum_{r \geq 0} h_{0,r}(t_0) t^r, \notag \\
 h_m(t,t_0)&=&\sum_{r \geq 0, r\neq 2m} h_{m,r}(t_0)t^r ,~\mbox{for}~m>0, \eea
and
\be h_{m,r}(t_0)=\sum_{s=0}^{\inf}h_{m,r,s}t_0^s. \ee
Now the convergent region for the expansion is
\be\label{regain1} |t_0|
<|t|
<|t_0|\cdot\left|\frac{e^{-2\pi TR}+e^{\pi T(u_1-u_2)}}{{e^{-2\pi TR}}-e^{\pi T(u_1-u_2)}}\right|. \ee
We can solve the wave function order by order with respect to $z_R$ and $t_0$.

To study the transformation between $\psi^{(3\pm)}$ and $\psi^{(2\pm)}$, we rewrite  $\psi^{(2\pm)}$ in terms of the  $t$ coordinate
\bea \psi^{(2\pm)}(t)&=&\psi^{(2\pm)}(z)(\frac{\partial z}{\partial t})^{-\frac{1}{2}} \notag \\
&=&(-2t_0)^{\frac{1}{2}}e^{\frac{1}{2}\pi T(u_1+u_2)}
\frac{1}{(1+t_0)^{\frac{1}{2}(1 \pm \frac{1}{n})}(1-t_0)^{\frac{1}{2}(1-\mp \frac{1}{n})}} \notag \\
&~&\cdot (1+t)^{\frac{1}{2}(1\pm \frac{1}{n})}(1-t)^{\frac{1}{2}(1\mp \frac{1}{n})}
g^{\pm}(z)\mid_{z=e^{\pi T(u_1+u_2)}\frac{t+t_0}{t-t_0}}, \eea
with convergent region
\be\label{regain2} \mid t \mid >t_0 \frac{e^{-2\pi TR}+e^{\pi T(u_2-u_1)}}{e^{\pi T(u_2-u_1)}-e^{-2\pi TR}}. \ee
 By comparing the coefficients in the expansions of two wave functions in the overlapping region
 \be t_0 \frac{e^{-2\pi TR}+e^{\pi T(u_2-u_1)}}{e^{\pi T(u_2-u_1)}-e^{-2\pi TR}}
<\mid t\mid
<|t_0|\cdot|\frac{e^{-2\pi TR}+e^{\pi T(u_1-u_2)}}{{e^{-2\pi TR}}-e^{\pi T(u_1-u_2)}}|,\ee
 we get
\bea
\left(
\begin{array}{ccc}
\psi^{(2-)} \\
\psi^{(2+)}
\end{array}
\right)
&\equiv&H_1 \cdot
\left( \begin{array}{ccc}
\psi^{(3-)} \\
\psi^{(3+)}
\end{array} \right),
\eea
where
\be
H_1=(-2t_0)^{\frac{1}{2}}e^{\frac{1}{2}\pi T(u_1+u_2)}
\left(
\begin{array}{ccc}
\frac{1}{(1+t_0)^{\frac{1}{2}(1-\frac{1}{n})}(1-t_0)^{\frac{1}{2}(1+\frac{1}{n})}}&0 \\
0& \frac{1}{(1+t_0)^{\frac{1}{2}(1+\frac{1}{n})}(1-t_0)^{\frac{1}{2}(1-\frac{1}{n})}}
\end{array}
\right)
\cdot\left(
\begin{array}{ccc}
T^{--}&T^{-+} \\
T^{+-}&T^{++}
\end{array}
\right).\ee
The explicit expression of the matrix elements of   $T$ are listed in Appendix F.

Similarly, we can rewrite the wave function $\psi^{(3)}$ in terms of  $u$ coordinate and compare the expansion coefficients of $\psi^{(3)}$ and $\psi^{(1)}$ in the region $z_R z_2<|z|<z_1$. We read the transformation
\bea
\left(
\begin{array}{ccc}
\psi^{(3-)} \\
\psi^{(3+)}
\end{array}
\right)
&\equiv&H_2 \cdot \left(
\begin{array}{ccc}
\psi^{(1-)} \\
\psi^{(1+)}
\end{array} \right).
\eea
where
\be
H_2= (-2t_0)^{\frac{1}{2}}
\left( \begin{array}{ccc}
(z_1z_2)^{\frac{1}{4n}}C^{-} &0 \\
0& (z_1z_2)^{-\frac{1}{4n}}C^{+}
\end{array} \right) .\ee
The perturbative expansions of the matrix elements $C^-$ and $C^+$ are listed in Appendix G.
Because  $\psi^{(1,+)}$ $\psi^{(3,+)}$ ($\psi^{(1,-)}$ $\psi^{(3,-)}$) share the same monodromy condition around the cycle encircling the origin, the transformation matrix $H_2$ is diagonal.

With these wave functions we can get the Schottky generators for the cycle $\tilde{A}^{(i)},i>1$. To transform the arguments between different sheets, we need another group element that denotes the action of circling around the branch point $u_1$ or the origin  counterclockwise. Under such action, the wave function  gets an extra phase so that the transformation matrix is
\bea M= \left( \begin{array}{ccc}
e^{-2\pi i \frac{1}{n}}& 0 \\
0& e^{2\pi i\frac{1}{n}}
\end{array} \right).
\eea

With these transformation elements, we can build the Schottky generators for the $\tilde{A}^{(i+1)}$ cycle as
\be L_i=M^{i-1}H_1M^{-1}H_1^{-1}M^{-i}, \ee
 with $i=1,...,n-1$. Ignoring the commutator in the fundamental group, they correspond to the thermal cycles in the $(i+1)$th sheet in the homology group.

The other Schottky generator corresponds to the horizontal cycle in the first sheet. To find the new generator, we need to discuss two other couples of the wave functions to cover the cycle.
Under a self-mapping conformal transformation
\be \tilde{z}=\frac{z_1 z_2}{z}, \ee
the energy momentum tensor does not change, so the wave functions  under the conformal transformation are
\be \psi^{(4,\pm)}(\tilde{z})=\psi^{(1,\pm)}(z)(\frac{\partial z}{\partial \tilde{z}})^{-\frac{1}{2}}
=(-1)^{-\frac{1}{2}}(z_1 z_2)^{\pm \frac{1}{2n}} \tilde{z}^{\frac{1}{2}(1\mp\frac{1}{n})}[1+\sum_{r \neq 0}
a_r\tilde{z}^r
], \ee
which is convergent in
\be z_2<|\tilde{z}|<\frac{z_1}{z_R}, \ee
and
\be\label{psi5} \psi^{(5,\pm)}(\tilde{z})=\psi^{(3,\pm)}(z)(\frac{\partial z}{\partial \tilde{z}})^{-\frac{1}{2}}
=(-1)^{-\frac{1}{2}}\psi^{(3,\mp)}(\tilde{z}), \ee
which is convergent at
\be |\tilde{z}-z_1|<\min \{z_1-z_2z_R,\frac{z_1}{z_R}-z_1\}. \ee

Taking a conformal transformation
\be\label{ztz} \tilde{z}=\frac{z}{z_R}, \ee
it is easy to see that the solutions $\psi^{(1)}$ and $\psi^{(4)}$ share the same convergent region,
and $\psi^{(1)}$ transforms as
\be \psi^{(1,\pm)}(\tilde{z})=\psi^{(1,\pm)}(z)(\frac{\partial z}{\partial \tilde{z}})^{-\frac{1}{2}}
z_R^{\pm\frac{1}{2n}}\tilde{z}^{\frac{1}{2}(1\pm\frac{1}{n})}[1+\sum_{r \neq 0}a_r \tilde{z}^r ].
\ee
Comparing the coefficients of the leading terms, we find
\bea \left( \begin{array}{ccc}
\psi^{(1,-)} \\
\psi^{(1,+)}
\end{array} \right)
= \left( \begin{array}{ccc}
z_R^{-\frac{1}{2n}}(-1)^{\frac{1}{2}}(z_1z_2)^{-\frac{1}{2n}}& 0 \\
0& z_R^{\frac{1}{2n}}(-1)^{\frac{1}{2}}(z_1z_2)^{\frac{1}{2n}}
\end{array} \right)
\left( \begin{array}{ccc}
0~1 \\
1~0
\end{array} \right)
\left( \begin{array}{ccc}
\psi^{(4,-)} \\
\psi^{(4,+)}
\end{array} \right).
\eea
The other wave functions $\psi^{(5)}$  are related to $\psi^{(3)}$ by the conformal transformation (\ref{ztz}) . Therefore we get the transformation \be
\left( \begin{array}{ccc}
\psi^{(4,-)} \\
\psi^{(4,+)}
\end{array} \right)
=H_2^{-1}
\left( \begin{array}{ccc}
\psi^{(5,-)} \\
\psi^{(5,+)}
\end{array}
\right).
\ee
Considering the relation (\ref{psi5}), we have the transformation
\bea
\left( \begin{array}{ccc}
\psi^{(5,-)} \\
\psi^{(5,+)}
\end{array} \right)
=(-1)^{-\frac{1}{2}} \left( \begin{array}{ccc}
0&1 \\
1&0
\end{array} \right)
\left( \begin{array}{ccc}
\psi^{(3,-)} \\
\psi^{(3,+)}
\end{array} \right).
\eea

%So that
%\be
%\left( \begin{array}{ccc}
%\psi^{(5,-)} \\
%\psi^{(5,+)}
%\end{array} \right)
%=(-2y_0)^{-\frac{1}{2}}
%\left( \begin{array}{ccc}
%u_R^{-\frac{1}{2n}}(C^{+})^{-1}(u_1u_2)^{-\frac{1}{4n}} &0 \\
%0& u_R^{\frac{1}{2n}}(C^{-})^{-1}(u_1u_2)^{\frac{1}{4n}}
%\end{array} \right)
%\left( \begin{array}{ccc}
%\psi^{(3,-)} \\
%\psi^{(3,+)}
%\end{array} \right)
%\ee
With these results, we obtain the generator
\bea L_n&=&H_1H_2
\left( \begin{array}{ccc}
z_R^{-\frac{1}{2n}}(-1)^{\frac{1}{2}}(z_1z_2)^{-\frac{1}{2n}}& 0 \\
0& z_R^{\frac{1}{2n}}(-1)^{\frac{1}{2}}(z_1z_2)^{\frac{1}{2n}}
\end{array} \right)
\left( \begin{array}{ccc}
0~1 \\
1~0
\end{array} \right)
H_2^{-1}
\left( \begin{array}{ccc}
0~1 \\
1~0
\end{array} \right)
H_1^{-1} \notag \\
&=&H_1 \left( \begin{array}{ccc}
(C^+)^{-1}C^{-}z_R^{-\frac{1}{2n}} &0 \\
0& C^+(C^{-})^{-1}z_R^{\frac{1}{2n}}
\end{array} \right) H_1^{-1}\equiv H_1NH_1^{-1}
\eea

Up to now, we have built all of the Schottky generators $L_i, i=1,\cdots n$. To calculate the 1-loop correction to the partition function, we need to find all of the primitive elements up to a conjugate. Even though there are infinite primitive conjugate classes,  only a finite number of them contribute at each order of the expansion with respect to $(u_1-u_2)$ and $z_R^{1/n}$. In this work, we are satisfied to  calculate the 1-loop correction up to order $(u_1-u_2)^4$ and $z_R^{3/n}$.

The Schottky group elements can be classified into two classes. In the first class,  the group elements are generated  by
\be\label{ged} L_{i}~ \mbox{with}~ i<n, \ee
and their inverses. They are similar to the ones in the double interval case as shown in \cite{Barrella:2013wja}. The simplest one is  \be \gamma_{i,k}=L_{i}L_{i+1}...L_{i+k}=M^{i-1}H_1M^{-(k+1)}H_1^{-1}M^{-(i+k)}. \ee
With this block,  all of the group elements in this class are generated by (\ref{ged}) and their inverses as
\be\label{group} \gamma_{k_1,k_2,...k_{2p},m}=M^{m}(\prod_{j=1}^p M^{k_{2j-1}}H_1M^{k_{2j}}H_1^{-1})M^{-m}. \ee
The other class involves  $L_n^{r}$, which can be written as
\bea L_n^{r}=H_1 N^r H_1^{-1}\sim H_1 M^{k^{'}} H_1^{-1}, \eea
with
\be\label{kp} k^{'}=\frac{n i}{2\pi} \log ((C^{+})^{-1}C^{-}z_R^{-\frac{1}{2n}}) \ee
As they take the similar form as (\ref{group}), all the elements take the general form of (\ref{group}).  However, in (\ref{group}),  there are nonprimitive elements,   and some of them are conjugate to each other. %We rewrite the elements in such a general form only to analyze the small interval properties.

 In the large interval limit, the asymptotic forms of the group elements are, respectively, 
\be H_1\sim (-2t_0)^{\frac{1}{2}}e^{\frac{1}{2}\pi T(u_1+u_2)}
\left( \begin{array}{ccc}
-\frac{n}{2t_0} ~ \frac{n}{2t_0} \\
-\frac{n}{2t_0}~ \frac{n}{2t_0}
\end{array}
\right),
\ee

\be H_1^{-1} \sim (-2t_0)^{-\frac{1}{2}}e^{-\frac{1}{2}\pi T(u_1+u_2)}
\left( \begin{array}{ccc}
-\frac{n}{2} ~ \frac{n}{2} \\
-\frac{n}{2} ~ \frac{n}{2}
\end{array}
\right),
\ee
Then a group element in (\ref{group} ) has an asymptotic form
\be \gamma_{k_1,k_2,...k_{2p},m} \sim
(\frac{n^2}{4t_0})^p \prod_{s=2}^{2p}(e^{\frac{2\pi i}{n}k_s}-e^{-\frac{2\pi i}{n}k_s})
\left( \begin{array}{ccc}
-e^{-\frac{2\pi i}{n}k_1}~e^{-\frac{2\pi i}{n}k_1}\\
-e^{\frac{2\pi i}{n}k_1}~e^{\frac{2\pi i}{n}k_1}
\end{array}
\right),
\ee
whose nonzero eigenvalue is
\bea q^{-\frac{1}{2}}=(\frac{n^2}{4t_0})^p \prod_{1=2}^{2p}(e^{\frac{2\pi i}{n}k_s}-e^{-\frac{2\pi i}{n}k_s})
(1+O(t_0)).\eea

Considering the large interval property for $t_0$
\be t_0\sim\frac{\pi T}{2}(u_2-u_1) ,\ee
we find that the 1-loop contribution to the partition function is
\be Z_q^{1-loop}=\sum_{s=2}^{\inf} \log (1-q^s)=q^2+q^3+O(q^4). \ee
As the leading order contribution to the 1-loop partition is at order $t_0^{4p}$,  we only need to consider the terms for $p \leq 1$, if we are only interested in the result up to order $l^4$. Furthermore in (\ref{group}), there may be some $k$'s that give the same terms as (\ref{kp}). Such terms are of order  $z_R^{-\frac{m}{n}}$, where $m$ is the number in (\ref{group}). Their leading contributions are of order  $z_R^{{4m}/{n}}$. If we only consider 1-loop contributions up to order $z_R^{3/n}$, we only need to consider $m \leq 1$.

Here we list the possible primitive conjugate classes, whose contributions to the 1-loop partition function are of order no higher than  $(u_1-u_2)^4$ or $z_R^{{3}/{n}}$.
\begin{enumerate}
\item The group element classes with no $L_n$, including $L_iL_{i+1}...L_{i+m-1}$ and their inverses. Their eigenvalues are
\be q_m^{(1)}=\frac{\pi^2 T^2}{4n^4}\frac{1}{\sin^4\frac{\pi m}{n}}(u_1-u_2)^2+O(u_1-u_2)^3 ,\ee
with degeneracy $2(n-m)$.
Their contributions to the partition function are
\bea Z^{(1-loop)}&=&\sum_{m=1}^{n-1} 2(n-m) \sum_{r=2}^{\inf}\log (1-(q_m^{(1)})^r) \notag \\
&=&\frac{\pi^4 T^4}{226800n^7}(n^2-1)(n^2+11)(3n^4+10n^2+227) (u_1-u_2)^4 +O(u_1-u_2)^5 \nn
\eea

\item   The elements $L_n$ and $L_n^{-1}$, both of which have eigenvalue
\be q^{(2)}=(C_+C_-^{-1})^2z_R^{{1}/{n}} .\ee
The resulting contributions are 
\bea
Z^{1-loop}&=&2\sum_{r=2}^{\inf}\log (1-(q^{(2)})^r) \notag \\
&=&2\{z_R^{{2}/{n}}[1+\frac{2(-1+n^2)}{3n^3}\pi^2T^2(u_1-u_2)^2 \notag \\
&~&-\frac{(-11-60n+15n^2+120n^3-9n^4-60n^5+5n^6)}{270n^7}\pi^4T^4(u_1-u_2)^4 \notag \\
&~&+O(z_R)+O((u_1-u_2)^5)] \notag \\
&~&+z_R^{{3}/{n}}[1+\frac{(-1+n^2)}{n^3}\pi^2T^2(u_1-u_2)^2 \notag \\
&~&-\frac{(-11-90n+15n^2+180n^3-9n^4-90n^5+5n^6)}{180n^7}\pi^4T^4(z_1-z_2)^4 \notag \\
&~&+O((u_1-u_2)^5)+O(z_R)]+O(z_R^{\frac{3}{n}}) \}
\eea
It is clear that the result is expanded with respect to $u_1-u_2$, $z_R$, and $z_R^{1/n}$.

\item The elements with all kinds of generators  include
\bea (L_1...L_{j_1})^{-1}L_n(L_1...L_{j_2})&=&M^{j_1+1}H_1M^{j_1}NM^{-j_2}H_1^{-1}M^{-j_2+1} \notag \\
&=&M^{j_1+1}(H_1M^{j_1-j_2}NH_1^{-1}M^{j_1-j_2}) M^{-(j_1+1)}, \eea
with $m=j_1-j_2 \neq 0$. Their eigenvalues are
\be q^{(3)}_m=-\frac{\pi^2 T^2(u_1-u_2)^2}{n^4}\frac{e^{\frac{2\pi im}{n}}}{\sin^2\frac{\pi m}{n}}(1+2e^{\frac{2\pi im}{n}}z_R^{{1}/{n}}+O(z_R)+O(z_R^{{1}/{n}})z_R^{{1}/{n}}, \ee
with degeneracy $2n$. The contributions to the partition function are, respectively,
\bea Z^{(1-loop)}_n&=&2n\sum_{m=1}^{n-1}\sum_{r=2}^{\inf} \log (1-(q^{(3)}_m)^r) \notag \\
&=&2n\{\frac{\pi^4 T^4}{n^8}(u_1-u_2)^4z_R^{{2}/{n}}\frac{1}{45}(-251+360n-110n^2+n^4) \\
&~&+\frac{\pi^4 T^4}{n^8}(u_1-u_2)^4 4z_R^{{3}/{n}}\frac{1}{45}(-1901+1440n-260n^2+n^4), \eea
for $n>3$, and
\be Z^{(1-loop)}_2=4(\frac{\pi^4 T^4 (u_1-u_2)^4}{2^8}(z_R-4z_R^{{3}/{2}}) \ee
for $n=2$, and
\be Z^{(1-loop)}_3=6(\frac{\pi^4 T^4(u_1-u_2)^4}{3^8 }(-\frac{16}{9}z_R^{{2}/{3}}+\frac{128}{9}z_R),
\ee
for $n=3$.
\end{enumerate}
%These results match with the CFT's calculation exactly.
Taking into account  all of the contributions, we obtain the 1-loop correction to the $n$th holographic R\'enyi entropy. For $n=2$, we have
\bea S_2^{1-loop}&=&-\frac{1}{2048}\pi^4 T^4(u_1-u_2)^4-\left(2+\frac{1}{2}\pi^2T^2(u_1-u_2)^2
+\frac{25}{384}\pi^4T^4(u_1-u_2)^4\right)z_R \notag \\
&~&-\left(2+\frac{3}{4}\pi^2T^2(u_1-u_2)^2+\frac{15}{256}\pi^4T^4(u_1-u_2)^4\right)z_R^{{3}/{2}}
+O((u_1-u_2)^5)+O(z_R^{2}). \notag \\ \eea
For $n=3$, we find
\bea S_3^{1-loop}&=&-\frac{16}{177147}\pi^4T^4(u_1-u_2)^4
-\left(1+\frac{16}{27}\pi^2T^2(u_1-u_2)^2+\frac{800}{59049}\pi^4T^4(u_1-u_2)^4\right)z_R^{{2}/{3}} \notag \\
&~&-\left(1+\frac{8}{27}\pi^2T^2(u_1-u_2)^2+\frac{280}{6561}\pi^4T^4(u_1-u_2)^4\right)z_R
+O((u_1-u_2)^5)+O(z_R^{{4}/{3}}). \notag \\ \eea
And for $n\geq 3$, we obtain
\bea S_n^{1-loop}&=&-\frac{\pi^4 T^4}{226800n^7}(n+1)(n^2+11)(3n^4+10n^2+227)(u_1-u_2)^4 \notag \\
&~&-\frac{1}{n-1}\left\{\left(2+\frac{4(n^2-1)}{3n^3}\pi^2T^2(u_1-u_2)^2 \right.\right.\notag \\
&~&\left.-\frac{299-444n+135n^2+24n^3-3n^4-12n^5+n^6}{27n^7} \pi^4T^4(u_1-u_2)^4\right)z_R^{{2}/{n} }\notag \\
&~&+\left(2+\frac{2(n^2-1)}{n^3}\pi^2T^2(u_1-u_2)^2 \right.\notag \\
&~&\left.\left. -\frac{6081-4626n+835n^2+36n^3-5n^4-18n^5+n^6}{18n^7}\pi^4T^4(u_1-u_2)^4\right)z_R^{{3}/{n}}\right\}\notag \\
&~&+O((u_1-u_2)^5)+O(z_R^{{4}/{n}})
\eea
For all the cases, the holographic results are in perfect match with the ones in the field theory up to the order we are interested in.

\section{Conclusion and discussion}

In this work, we completed our study on the R\'enyi entropy of  a large interval on a torus in the light of AdS$_3$/CFT$_2$ correspondence. In the case that the interval is not so large, we may expand the density matrix in the CFT level by level and compute the entropy perturbatively; while on the bulk side, we can follow the prescription in \cite{Barrella:2013wja} and take into account the size dependence \cite{Chen:2014unl} to read the holographic R\'enyi entropy, which is in good agreement with the CFT computation. However, when the interval is large, the problem becomes quite difficult. On the field side, the perturbative prescription used in the short interval case breaks down, and we have to find another effective way to compute the partition function. On the bulk side, the dual gravitational configurations are different from the ones in the short interval case, as indicated in the study in \cite{Azeyanagi:2007bj}.

To overcome these difficulties, we developed a new prescription and treatment in both field theory and dual gravity. On the field theory side,
we proposed in \cite{Chen:2014hta} to insert a complete set of state bases in the twist sector of orbifold CFT
to compute the large interval R\'enyi entropy. We applied this proposal in this paper and focused on the vacuum module of the CFT dual to the pure AdS$_3$ gravity.
We found that the leading linear $c$ contributions were dominated by the twist vacuum module and the subleading ones got contributions from all the twist states. This allows us to read the leading contributions by applying the Ward identity to the correlation function of four twist operators, two at the branch points and the other two at the left and the right infinities of the cylinder. We did find the holographic entanglement entropy suggested in \cite{Azeyanagi:2007bj}.

On the gravity side, we suggested a new set of monodromy conditions on the cycles to construct the Schottky generators and corresponding gravitational configurations. To check the validness of the monodromy condition, we computed the holographic entanglement entropy and reproduced successfully the expected value. We read the classical part of the holographic R\'enyi entropy  by integrating two differential equations (\ref{partialm}) and (\ref{partialc}), one encoding the dependence of HRE on the moduli parameter of the Schottky space and the other on the size of the torus. Moreover we discussed carefully the 1-loop correction to the HRE, following the treatment in \cite{Barrella:2013wja}. We found good agreements of classical contribution and 1-loop quantum correction to the HRE with the leading and subleading large $c$ results in the field theory, up to the first few orders.
For the classical part, the agreement is up to $e^{-4\pi TR}$ and $l^6$ orders, while for the quantum part, the agreement is up to $e^{-\frac{6\pi TR}{n}}$ and $l^4$ orders. %(more discussion on the difference of the expansion on two sides)

The study in this work presents another piece of evidence to strongly support the holographic computation of the entanglement entropy in the context of the AdS$_3$/CFT$_2$ correspondence. Taking into account the accumulated evidence on the holographic R\'enyi entropy in the cases including double-interval and single short interval on the torus, it suggests that the holographic computation is exact perturbatively not only at the classical level but also at  the1-loop quantum level. Furthermore, our field theory study shows that there are actually $1/c$ corrections in the partition function when the Riemann surface is of higher genus than $1$. It would be interesting to see if the agreement could go beyond the 1-loop level\cite{Yin:2007gv,Headrick:2015gba} or even nonperturbatively.

Our study could be generalized to other cases. In particular, it is interesting to study the higher spin R\'enyi entropy of the single interval on a torus by direct field theory computation \cite{Datta:2014ska,Datta:2014uxa,Long:2014oxa,Datta:2014zpa,deBoer:2014sna}and Wilson line prescription in the bulk\cite{Ammon:2013hba,deBoer:2013vca,Castro:2014mza}.

The study of holographic entanglement entropy may shed light on the AdS$_3$ quantum gravity\cite{Witten:2007kt,Maloney:2007ud}. There are two essential questions on the quantum AdS$_3$ gravity. One is on the precise definition of the quantum gravity,  string theory, or something else. The other is on the construction of the dual CFT.  There is ample evidence, for example, the work in \cite{Maloney:2007ud}, that the dual CFT might not exist.  However, the results in this work and other related ones suggest that 
there exists an equivalence between the semiclassical AdS$_3$ gravity including the pure gravity sector and  the large central charge limit of a 2D CFT, 
which has a sparse light spectrum\cite{Hartman:2013mia,Hartman:2014oaa,Keller:2014xba,Haehl:2014yla,Belin:2014fna}. In our study, it turns out that in the large central charge limit,  the vacuum module dominates the contribution to the partition function. It is not clear when the states in the sparse light spectrum begin to contribute. Moreover, the regulated on-shell action of the gravitational configuration in the AdS$_3$ gravity is a Liouville theory. This raises the issue if the dual CFT could be a Liouville CFT. For a recent study on this issue, see \cite{Jackson:2014nla}. It would be interesting to see if it is possible to prove the equivalence by using the Liouville theory.\footnote{We would like to thank the anonymous referee for pointing out this possibility.}

\vspace*{10mm}
\noindent {\large{\bf Acknowledgments}}\\
J.Q would like to thank the participants of the workshop ``International Workshop on Condensed Matter Physics and AdS/CFT" for comments on the work, and the organizers of the workshop for hospitality.  The work was  supported in part by NSFC Grants No.~11275010, No.~11335012 and No.~11325522.
\vspace*{5mm}

\begin{appendix}

\section{States in the vacuum module}
In this section, we list some low-lying excited  states in the vacuum module. We focus only on the holomorphic sector. For the antiholomorphic sector, it is similar to the holomorphic one. The first few excited states up to level 4 are, respectively, 
\bea \mid 2\rangle&=&\sqrt{\frac{2}{c}}L_{-2}\mid 0\rangle \notag \\
\mid 3\rangle&=&\sqrt{\frac{1}{2c}}L_{-3}\mid 0\rangle \notag \\
\mid 4,1\rangle &=&\sqrt{\frac{1}{5c}}L_{-4}\mid 0\rangle \notag \\
\mid 4,2\rangle &=&(\frac{c^2}{2}+\frac{11}{5}c)^{-\frac{1}{2}}(L_{-2}L_{-2}-\frac{3}{5}L_{-4})\mid 0\rangle. \eea
The corresponding vertex operators at the origin and the infinity take the forms, respectively,
\bea L_{-2}&\rightarrow& T(w)\mid_{w=0} \notag \\
&\rightarrow& w^4T(w)\mid_{w\rightarrow \inf} \notag \\
L_{-3} &\rightarrow& \partial T(w)\mid_{w=0} \notag \\
&\rightarrow & -w^6\partial T(w)-4w^5T(w)\mid_{w\rightarrow \inf}  \notag \\
L_{-4} &\rightarrow & \frac{1}{2}\partial^2 T(w)\mid_{w=0} \notag \\
&\rightarrow &
\frac{1}{2}w^8\partial^2T(w)+5w^7\partial T(w)+10w^6T(w)\mid_{w\rightarrow \inf} \notag \\
L_{-2}L_{-2}-\frac{3}{5}L_{-4}
&\rightarrow &
:T(w)^2:-\frac{3}{10}\partial^2T(w)\mid_{w=0} \notag \\
&\rightarrow&  w^8(:T(w)^2:-\frac{3}{10}\partial^2T(w))\mid_{w\rightarrow \inf}\notag
\eea

\section{Conformal transformation for $:T^2(z):$}

In the calculation, we need the conformal transformation of $:T^2(z):$, which is not a primary operator. Under a conformal transformation $z\to w(z)$, we have
\be T(z)=T(w)(\frac{\partial w}{\partial z})^2+\frac{c}{12}\{w,z\}, \ee
where
\be \{w,z\}=\frac{w'''}{w'}
\frac{3}{2}(\frac{w''}{w'})^2 \ee
is the Schwarzian derivative and the prime denotes the derivative with respect to $z$.
For $:T^2(z):$, we have
\bea :T^2(z):&=&\oint\frac{dz_1}{z_1-z}T(z_1)T(z) \notag \\
&=&\frac{c}{1440}(-225(\frac{w''}{w'})^4
+480(\frac{w''}{w'})^2
\frac{w'''}{w'}
-100(\frac{w'''}{w'})^2
-180\frac{w''}{w'}
\frac{w''''}{w'} \notag \\
&~&+36\frac{w'''''}{w'})
+2T(w)(w')^2
(-\frac{1}{4}(\frac{w''}{w'})^2
+\frac{2}{3}\frac{w'''}{w'})
+\partial T(w)(w')^3\frac{3}{2}
\frac{w''}{w'}  \notag \\
&~&+:T^2(w):(w')^4+T(w)(w')^2\frac{c}{6}\{w,z\}
+(\frac{c}{12}\{w,z\})^2 .
\eea
Especially for the conformal transformation
\be w=e^{\frac{2\pi}{n\beta}y}, \ee
we have
\be T(y)=T(w)(\frac{\partial w}{\partial y})^2-\frac{\pi^2}{6n^2\beta^2}c, \ee
\be :T(y)^2:=(\frac{c^2}{36}+\frac{11c}{90})\frac{\pi^4}{n^4\beta^4}
+(-\frac{c}{3}+\frac{10}{3})T(w)(\frac{\partial w}{\partial y})^2\frac{\pi^2}{n^2\beta^2}
+3\partial T(w)(\frac{\partial w}{\partial y})^3\frac{\pi}{n\beta}+:T(w)^2:(\frac{\partial w}{\partial y})^4 .\notag\ee

\section{Correlation functions $f_i$}

In this computation of the R\'enyi entropy by inserting the twist sector states, we need to compute the correlation functions $f_i$. Here we list the results for the first few ones  needed in the relation (\ref{Snt}).
\bea \log \langle t\mid V(0)\mid t\rangle
&=&c(-\frac{(-1 + n^2)\pi^2}{72\beta^2n^3}l^2+\frac{(-11+20n^2-18 n^4+8n^6+n^8)\pi^4}{25920\beta^4n^7}l^4) \notag \\
&~&+\frac{(-2497+2160n^2+294n^4+40n^6+3n^8)\pi^4}{453600 \beta^4 n^7}l^4+O(l^5), \eea
\bea f_2&=&(1+\frac{2(-1+n^2)\pi^2}{3\beta^2n^3}l^2
-\frac{(299-444n+135n^2+24n^3-3n^4-12n^5+n^6)\pi^4}{54\beta^4n^7}l^4)  \notag \\
&~&-\frac{1}{c}\frac{(733-672n^2-63n^4+2n^6)\pi^4}{945\beta^4n^7}l^4+O(l^5) ~~~\mbox{for}~ n\neq1,2 \notag  \\
&=&1~~~\mbox{for}~n=1 \notag \\
&=&c\frac{\pi^4}{128\beta^4}l^4+(1+\frac{\pi^2}{4\beta^2}l^2+\frac{25\pi^4}{768\beta^4}l^4)
 +\frac{1}{c}\frac{3\pi^4}{128\beta^4}l^4+O(l^5) ~~~\mbox{for}~n=2,
\eea

\bea f_3&=&
(1+\frac{(-1+n^2)\pi^2}{\beta^2n^3}l^2-\frac{(6081-4626n+835n^2+36n^3-5n^4-18n^5+n^6)\pi^4}{36\beta^4n^7}l^4) \notag \\
&~&  -\frac{1}{c}\frac{(-933+1148n^2-217n^4+2n^6)\pi^4}{630\beta^4n^7}+O(l^5)
   ~~~\mbox{for}~n\neq1,3 \notag \\
&~&1~~~\mbox{for}~n=1 \notag \\
&~&c\frac{128\pi^4}{6561\beta^4}l^4+(1+\frac{8\pi^2}{27\beta^2}l^2+\frac{280\pi^4}{6561\beta^4}l^4)
+\frac{1}{c}\frac{32\pi^4}{6561\beta^4}l^4+O(l^5)~~~\mbox{for}~n=3 ,\eea

\bea f_{4,1}&=&
(1+\frac{4(-1+n^2)\pi^2}{3\beta^2n^3}l^2
-\frac{(201919-108120n+14115n^2+240n^3-39n^4-120n^5+5n^6)\pi^4}{135\beta^4n^7}l^4) \notag \\
&~&-\frac{1}{c}\frac{2(83213-136080n+55020n^2-2163n^4+10n^6)\pi^4}{4725\beta^4n^7}l^4+O(l^5)
~~~\mbox{for}~n\neq1,2,4 \notag \\
&~&1~~~\mbox{for}~n=1 \notag  \\
&~&c\frac{5\pi^4}{64\beta^4}l^4+(1+\frac{\pi^2}{2\beta^2}l^2+\frac{473\pi^4}{1920\beta^4}l^4)
+\frac{1}{c}\frac{3\pi^4}{320\beta^4}l^4+O(l^5)~~~\mbox{for}~n=2 \notag \\
&~&c\frac{125\pi^4}{4096\beta^4}l^4+(1+\frac{5\pi^2}{16\beta^2}l^2+\frac{2261\pi^4}{49152\beta^4}l^4)
+\frac{1}{c}\frac{99\pi^4}{40960\beta^4}l^4+O(l^5)~~~\mbox{for}~n=4 ,
\eea

\bea f_{4,2}&=&
(1+\frac{4(-1+n^2)\pi^2}{3\beta^2n^3}l^2
-\frac{(299-456n+135n^2+48n^3-3n^4-24n^5+n^6)\pi^4}{27\beta^4n^7}l^4) \notag \\
&~&-\frac{1}{c}\frac{2(124247-166320n+43260n^2-1197n^4+10n^6)\pi^4}{4725\beta^4n^7}l^4+O(l^5)
~~~\mbox{for}~n\neq1,2 \notag \\
&~&1~~~\mbox{for}~n=1 \notag \\
&~&c\frac{\pi^4}{64\beta^4}l^4+(1+\frac{\pi^2}{2\beta^2}l^2+\frac{337\pi^4}{1920\beta^4}l^4)+
\frac{1}{c}\frac{57\pi^4}{320\beta^4}l^4~~~\mbox{for}~n=2 .
\eea

%\section{Canonical cycles}

%In this section, we list the canonical cycles for the Riemann surface in the double interval case and single interval finite temperature case. In the field theory, we insert a complete basis along $a_i$ cycles. In the holographic calculation, we take a Schottky uniformization for the Riemann surface: $a_i$ cycles are trivial cycles and $b_i$ cycles correspond to Schottky generators.
%\begin{figure}\label{cycle}
%\centering
%\subfloat[Canonical cycles for double interval]{\includegraphics[width=6cm]{cycle1.jpg}\label{cycle1}}
%\quad
%\subfloat[Canonical cycles for single interval finite %temperature]{\includegraphics[width=6cm]{cycle2.jpg}\label{cycle2}}
%\\
%\caption{The canonical cycles for the double interval case and the single interval finite temperature case. The double line denote the branch cut for $n=3$. The solid line denote $a_i$ cycles and the dashed line denote $b_i$ cycles. The canonical cycles satisfy the intersection relation $a_i\circ b_j=\delta_{i,j}$ \cite{Dijkgraaf:1987vp}.}
%\end{figure}

\section{Mutual R\'enyi information for the double intervals}

In this appendix, we list the R\'enyi mutual information for the double intervals, which has been computed in \cite{Chen:2013kpa}. In terms of a small cross ratio $x$, the leading and next-to-leading contributions are, respectively,
. \bea \lefteqn{I_n^{tree}(x,\bar{x})=} \notag \\
&&\frac{c(n-1)(n+1)^2x^2}{288n^3}+\frac{c(n-1)(n+1)^2x^3}{288n^3}+
\frac{c(n-1)(n+1)^2(1309n^4-2n^2-11)x^4}{414720n^7} \notag \\
&~&+\frac{c(n-1)(n+1)^2(589n^4-2n^2-11)x^5}{207360n^7} \notag  \\
&~&+\frac{c(n-1)(n+1)^2(805139n^8-4244n^6-23397n^4-86n^2+188)x^6}{313528320n^{11}}+O(x^7)+c.c. \eea

\bea \lefteqn{I_n^{1-loop}(x,\bar{x})} \notag \\
&=&\frac{(n+1)(n^2+11)(3n^4+10n^2+227)x^4}{7257600n^7} \notag  \\
&~&+\frac{(n+1)(109n^8+1495n^6+11307n^4+81905n^2-8416)x^5}{119750400n^9} \notag  \\
&~&+\frac{(n+1)(1444050n^{10}+19112974n^8+140565305n^6+1000527837n^4-167731255n^2-14142911)x^6}{1046139494400n^{11}}
\notag  \\
&~&+O(x^{7})+c.c.
\eea

\section{The accessory parameters}

After imposing the monodromy condition, the accessory parameters $\gamma$ and $\delta$ can be solved order by order.
 Here we just list the expansion coefficients of the first
few orders
\be \gamma_{0,-1}=-\frac{2 \pi  (n-1) (n+1) T z_2}{n^2} ,\hs{3ex} \gamma_{0,0}=-\frac{\pi  (n-1) (n+1) T}{n^2}, \nn\ee
\be \gamma _{0,1}=\frac{\pi  (n-1)^2 (n+1)^2 T}{6 n^4 z_2} ,\hs{3ex}
 \gamma_{0,2}=-\frac{(-1 + n)^2 (1 + n)^2 \pi T}{12 n^4 z_2^2},\nn\ee
\be \gamma_{0,3}=\frac{\pi  (n-1)^2 (n+1)^2 \left(229 n^4-2 n^2-11\right) T}{4320 n^8 z_2^3} ,\nn\ee
\be \gamma_{0,4}=-\frac{\pi  (n-1)^2 (n+1)^2 \left(109 n^4-2 n^2-11\right) T}{2880 n^8 z_2^4} ,\nn\ee
\be \gamma_{0,5}=\frac{\pi  (n-1)^2 (n+1)^2 \left(62999 n^8-1724 n^6-9537 n^4-86 n^2+188\right) T}{2177280 n^{12} z_2^5}, \nn \ee
\be \gamma_{0,6}=-\frac{\pi  (n-1)^2 (n+1)^2 \left(20159 n^8-716 n^6-3993 n^4-86 n^2+188\right) T}{870912 n^{12} z_2^6}, \nn\ee
\bea \gamma_{0,7}&=&\frac{\pi  (n-1)^2 (n+1)^2 T}{15676416000 n^{16} z_2^7}\left(299803787 n^{12} \right.\nn\\
& &\left. -12840306 n^{10}-72253947 n^8-2555548 n^6+5564373 n^4+67854 n^2-58213\right) \nn\eea

\be \gamma_{1,0}= 0,\hs{3ex} \gamma_{1,1}= 0 ,\hs{3ex}\gamma_{1,2}= 0, \nn\ee
\be \gamma_{1,3}= \frac{\pi  (n-1)^3 (n+1)^3 T}{18 n^6 z_2^3} ,\hs{3ex} \gamma_{1,4}= -\frac{\pi  (n-1)^3 (n+1)^3 T}{12 n^6 z_2^4} ,\nn\ee
\be \gamma_{1,5}= \frac{\pi  (n-1)^3 (n+1)^3 \left(439 n^4-2 n^2-11\right) T}{4320 n^{10} z_2^5} ,\nn \ee
\be \gamma_{1,6}= -\frac{\pi  (n+1)^3 \left(199 n^7-597 n^6+595 n^5-193 n^4-17 n^3+35 n^2-33 n+11\right) T}{1728 n^{10} z_2^6} , \nn\ee
\bea \gamma_{1,7}&=& \frac{\pi  (n+1)^3  T}{32659200 n^{14} z_2^7} \left(4110527 n^{11}-12331581 n^{10}+12268009 n^9-3919811 n^8\right. \nn\\
 & &\left. -541434 n^7+1115726 n^6-1053566 n^5+354954 n^4+371 n^3-12409 n^2+13821 n-4607\right)\nn\eea

\be \gamma_{2,0}= 0 ,\hs{3ex} \gamma_{2,1}= 0 ,\hs{3ex}\gamma_{2,2}= 0, \nn\ee
\be \gamma_{2,3}= \frac{\pi  (n-1)^2 (n+1)^2 \left(3 n^2-1\right) T}{6 n^6 z_2^3},\nn \ee
\be \gamma_{2,4}= -\frac{\pi  (n-1)^2 (n+1)^2 \left(3 n^2-1\right) T}{4 n^6 z_2^4} ,\nn\ee
\be \gamma_{2,5}= \frac{\pi  (n-1)^2 (n+1)^2 \left(1469 n^6-521 n^4+n^2+11\right) T}{1440 n^{10} z_2^5} ,\nn\ee
\be \gamma_{2,6}= -\frac{\pi  (n-1)^2 (n+1)^2 \left(749 n^6-281 n^4+n^2+11\right) T}{576 n^{10} z_2^6}, \nn \ee
\bea \gamma_{2,7}&=& \frac{\pi  (n+1)^2 T}{10886400 n^{14} z_2^7} \left(17287061 n^{12}-34574122 n^{11}+10493170 n^{10}+13587782 n^9 -6779897 n^8\right. \nn\\
 & & \left. -27988 n^7+404292 n^6-780596 n^5+383843 n^4+12910 n^3-11062 n^2+9214 n-4607\right) \nn\eea

\be \delta_0=-\frac{2\pi^2T^2}{n^2} ,\hs{3ex} \delta_{1,0}=\delta_{1,1}=\delta_{1,2}=\delta_{1,3}=0 ,\nn\ee
\be \delta_{1,4}=\frac{\pi ^2 \left(n^2-1\right)^3 T^2}{36 n^6 z_2^4} ,\hs{3ex}\delta_{1,5}=-\frac{\pi ^2 (n-1)^3 (n+1)^3 T^2}{18 n^6 z_2^5}, \nn \ee
\be \delta_{1,6}=\frac{\pi^2\left(n^2-1\right)^3 \left(1039  n^4 -2  n^2 -11  \right)T^2}{12960 n^{10}  z_2^6} , \nn\ee
\be \delta_{1,7}=-\frac{\pi^2\left(n^2-1\right)^3 \left(439  n^4 -2  n^2 -11 \right)T^2}{4320 n^{10} z_2^7},\nn \ee
\be \delta_{1,8}=\frac{\pi ^2 (n-1)^3 (n+1)^3 \left(15726467 n^8-116492 n^6-641778 n^4-1412 n^2+4607\right) T^2}{130636800 n^{14} z_2^8} ,\nn \ee
\be \delta_{1,9}=-\frac{\pi ^2 (n-1)^3 (n+1)^3 \left(4472147 n^8-45932 n^6-253698 n^4-1412 n^2+4607\right) T^2}{32659200 n^{14} z_2^9}, \nn \ee
\bea \delta_{1,10}&=&\frac{\pi^2(n-1)^3 (n+1)^3 T^2}{47029248000 n^{18} z_2^{10}} \left(7132759447  n^{12} -93106446 n^{10}  \right. \nn\\
& & \left. -515825187  n^8 -4929908  n^6 +16057833  n^4 +73554  n^2 -78893  \right) \nn \eea

\be \delta_{2,0}=\delta_{2,1}=\delta_{2,2}=\delta_{2,3}=0, \nn \ee
\be \delta_{2,4}=\frac{\pi ^2 \left(n^2-1\right)^2 \left(3 n^2-1\right) T^2}{6 n^6 z_2^4}, \nn\ee
\be \delta_{2,5}=-\frac{\pi ^2 \left(n^2-1\right)^2 \left(3 n^2-1\right) T^2}{3 n^6 z_2^5},\nn \ee
\be \delta_{2,6}=\frac{\pi ^2 \left(n^2-1\right)^2 \left(3269 n^6-1121 n^4+n^2+11\right) T^2}{2160 n^{10} z_2^6} ,\nn \ee
\be \delta_{2,7} =-\frac{\pi ^2 \left(n^2-1\right)^2 \left(1469 n^6-521 n^4+n^2+11\right) T^2}{720 n^{10} z_2^7}, \nn \ee
\be \delta_{2,8}=\frac{\pi ^2 \left(n^2-1\right)^2 \left(56156801 n^{10}-20579551 n^8+40454 n^6+681358 n^4-6455 n^2-4607\right) T^2}{21772800 n^{14} z_2^8},\nn \ee
\be \delta_{2,9}=-\frac{\pi ^2 \left(n^2-1\right)^2 \left(17031281 n^{10}-6432271 n^8+5174 n^6+293278 n^4-6455 n^2-4607\right) T^2}{5443200 n^{14} z_2^9},\nn \ee
\bea \delta_{2,10}&=&\frac{\pi ^2 \left(n^2-1\right)^2T^2}{7838208000 n^{18} z_2^{10}}  \left(28908872157 n^{14}-11219722453 n^{12}\right. \nn\\
& &\left. -16128831 n^{10}+654870399 n^8-21992209 n^6-17014839 n^4+188883 n^2+78893\right)\nn \eea

Collecting all these coefficients and changing back to the $u$ coordinate, we find
\bea \gamma_0&=&(\frac{1}{n^2}-1)\frac{1}{u_1-u_2}+\left(n^2-1\right) \pi ^2 T^2\left\{(-\frac{1}{3 n^4})(u_1-u_2) \right.\notag \\
&~&+(\frac{\pi ^2 T^2\left(n^6+9 n^4-9 n^2+11\right) }{540 n^8})(u_1-u_2)^3  \notag \\
&~&+(-\frac{\pi ^4 T^4\left(19 n^{10}+85 n^8-125 n^6+251 n^4-274 n^2+188\right) }{68040 n^{12}})(u_1-u_2)^5 \notag \\
&~&+\frac{\pi ^6T^6}{122472000 n^{16}} \left(4187 n^{14}+12787 n^{12}-22521 n^{10}+55919 n^8-100079 n^6
\right. \notag \\
&~&\left.\left.+143481 n^4-126067 n^2+58213\right) (u_1-u_2)^7+O((u_1-u_2)^8)\right\}\notag
\eea

\bea \gamma_{1}&=&\pi ^4  T^4(n^2-1)^3 \left\{\frac{4 }{9 n^6}(u_1-u_2)^3+
\frac{\pi ^2T^2 \left(19 n^4-2 n^2-11\right) }{135 n^{10}}(u_1-u_2)^5+ \right.\notag \\
&~&\left. \frac{\pi ^4T^4 \left(4187 n^8-572 n^6-4218 n^4-1412 n^2+4607\right) }{255150 n^{14}}
(u_1-u_2)^7+O((u_1-u_2)^8) \right\} \notag
\eea

\bea \gamma_{2}&=&\pi ^4T^4(n^2-1)^2\left\{ \frac{4   \left(3 n^2-1\right) }{3 n^6}(u_1-u_2)^3 \right. \notag \\
&~&+\frac{\pi ^2T^2 \left(209 n^6-101 n^4+n^2+11\right) }{45 n^{10}}(u_1-u_2)^5 \notag \\
&~&+\frac{\pi ^4T^4 \left(180041 n^{10}-104551 n^8-17506 n^6+43798 n^4-6455 n^2-4607\right) }{85050 n^{14}}(u_1-u_2)^7 \notag \\
&~&\left.+O((u_1-u_2)^8)\right\}\notag
 \eea

\bea \delta_1&=&\pi ^6 T^6(n^2-1)^3 \left\{ \frac{4 }{9 n^6}(u_1-u_2)^4
+\frac{2 \pi ^2T^2 \left(19 n^4-2 n^2-11\right) }{405 n^{10}}(u_1-u_2)^6 \right.\notag \\
&~&+\frac{\pi ^{4} T^4 \left(4187 n^8-572 n^6-4218 n^4-1412 n^2+4607\right) }{510300 n^{14}}
(u_1-u_2)^8  \notag \\
&~&\left.+O((u_1-u_2)^9)\right\}\notag\eea

\bea \delta_2&=&\pi ^6 T^6 (n^2-1)^2\left\{ \frac{8  \left(3 n^2-1\right)}{3 n^6}(u_1-u_2)^4\right. \notag \\
&~&+\frac{4 \pi ^2 T^2\left(209 n^6-101 n^4+n^2+11\right) }{135 n^{10}} (u_1-u_2)^6 \notag \\
&~&+\frac{\pi ^4T^4 \left(180041 n^{10}-104551 n^8-17506 n^6+43798 n^4-6455 n^2-4607\right) }{85050 n^{14}}(u_1-u_2)^8 \notag \\
&~&\left.+O((u_1-u_2)^8)\right\}\notag
\eea

\section{T matrix}

In this section, we list the leading order terms in the expansion of the $T$ matrix in the Schottky transformation. For any matrix element, we may expand it as
\be T=\sum_{r=0}^{\inf}\sum_{n=-1}^{\inf} T_{r,n}t_0^n z_R^r \ee
There are relations among the matrix elements
\be T^{-+}_{r,n}=(-1)^nT^{--}_{r,n} \ee
\be T^{++}=T^{--}|_{n \rightarrow -n} \ee
\be T^{+-}=T^{-+}|_{n \rightarrow -n} \ee
For the matrix element $T^{--}$, its expansion coefficients are, respectively,
\be T^{--}_{0,-1}=-\frac{n}{2} ,\hs{3ex}T^{--}_{0,0}=-\frac{1}{2n} \notag\ee
\be T^{--}_{0,1}=-\frac{n^4-6 n^3-5 n^2+6 n+4}{36 n^3} \notag\ee
\be T^{--}_{0,2}=\frac{-5 n^4+6 n^3+n^2-6 n+4}{36 n^5} \notag\ee
\be T^{--}_{0,3}=\frac{-547 n^8+3300 n^7+1450 n^6-4320 n^5+489 n^4+900 n^3-2600 n^2+120 n+1208}{32400 n^7} \notag\ee
\be T^{--}_{0,4}=-\frac{2563 n^8-4500 n^7+230 n^6+4320 n^5-4521 n^4+2700 n^3-280 n^2-2520 n+2008}{32400 n^9} \notag\ee
\bea T^{--}_{0,5}&=&-\frac{1}{28576800 n^{11}} (351301 n^{12}-2154138 n^{11}-482727 n^{10}+3021438 n^9-1320165 n^8-443814 n^7 \notag \\
&~&+2165179 n^6-767046 n^5 +202692 n^4+325752 n^3-1518552 n^2+17808 n+602272) \notag\eea
\bea T^{--}_{0,6}&=&\frac{1}{28576800 n^{13}}
(-1573097 n^{12}+3334842 n^{11}-636405 n^{10}-3026142 n^9+4008417 n^8-3258234 n^7 \notag \\
&~&+929665 n^6+1999494 n^5-2853228 n^4+2430792 n^3-1107960 n^2-1480752 n+1232608) \notag\eea
\bea T^{--}_{0,7}&=&\frac{1}{51438240000 n^{15}}
(-500204179 n^{16}+3117021000 n^{15}+239603780 n^{14}-4477586640 n^{13} \notag \\
&~&+2898902806 n^{12} +236602800 n^{11}-3339143740 n^{10}+1922809920 n^9-1609248619 n^8 \notag \\
&~&-247284600 n^7+2744490640 n^6-889423920 n^5+1223043536 n^4+325780800 n^3-2411735680 n^2 \notag \\
&~&+12080640 n+754291456)\notag
\eea
\be T^{--}_{1,-1}=T^{--}_{1,0}=T^{--}_{1,2}=0 , \hs{3ex} T^{--}_{1,1}=\frac{4 \left(n^2-1\right)}{3 n} \notag\ee
\be T^{--}_{1,3}=(4 (12 + 5 n - 23 n^2 - 10 n^3 - 2 n^4 + 5 n^5 + 13 n^6))/(45 n^5) \notag\ee
\be T^{--}_{1,4}=-\frac{2 \left(13 n^6-12 n^5-24 n^4+24 n^3+9 n^2-12 n+2\right)}{27 n^7} \notag \ee
\bea T^{--}_{1,5}&=&-\frac{1}{85050 n^9}
(-101693 n^{10}-116760 n^9-33317 n^8+275520 n^7+267361 n^6-201600 n^5 \notag \\
&~&-58831 n^4+43680 n^3-130168 n^2-840 n+56648)\notag
\eea
\be T^{--}_{2,-1}=T^{--}_{2,0}=T^{--}_{2,2}=0 ,\hs{3ex}T^{--}_{2,1}=\frac{4 \left(n^2-1\right)}{n} \notag\ee
\be T^{--}_{2,3}=\frac{4 \left(62 n^6+75 n^5+14 n^4-90 n^3-112 n^2+15 n+36\right)}{45 n^5} \notag \ee

\section{ C matrix}

For the matrix elements $C^+$ and $C^-$, their expansions are similar
\be C^{-}=\sum_{r=0}^{\inf}\sum_{n=0}^{\inf} C^{-}_{r,n} z_R^r t_0^n \ee
and
\be C^{+}=C^{-}\mid _{n \rightarrow -n} \ee
Here we list the ones for $C^-$:
\be C^{-}_{0,1}=C^{-}_{0,3}=C^{-}_{0,5}=C^{-}_{0,7}=0 \notag\ee
\be C^{-}_{0,0}=1 , \hs{3ex}C^{-}_{0,2}=\frac{-n^3-2 n^2+n+2}{6 n^3}\notag \ee
\be C^{-}_{0,4}=\frac{-67 n^7-140 n^6+114 n^5+48 n^4-63 n^3+180 n^2+16 n-88}{1080 n^7}\notag \ee
\bea C^{-}_{0,6}&=&\frac{1}{136080 n^{11}}(-4919 n^{11}-10570 n^{10}+10681 n^9-322 n^8-6053 n^7+17546 n^6 \notag \\
&~&-2893 n^5+2506 n^4+3872 n^3-15176 n^2-688 n+6016) \notag\eea
\bea C^{-}_{0,8}&=&\frac{1}{244944000 n^{15}}
(-6094181 n^{15}-13347880 n^{14}+15437132 n^{13}-4050896 n^{12}-8054742 n^{11} \notag \\
&~&+24501840 n^{10}-8606276 n^9+13161728 n^8+5132627 n^7-23387240 n^6+5905056 n^5-13379568 n^4 \notag \\
&~&-4416704 n^3+23953280 n^2+697088 n-7451264)\notag
\eea
\be C^{-}_{1,0}=C^{-}_{1,1}=C^{-}_{1,3}=C^{-}_{1,5}=0, \hs{3ex} \ C^{-}_{1,2}=\frac{4 \left(n^2-1\right)}{3 n^2} \notag\ee
\be C^{-}_{1,4}=\frac{2 \left(83 n^6-90 n^5-42 n^4+180 n^3-93 n^2-90 n+52\right)}{135 n^6} \notag \ee
\bea C^{-}_{1,6}&=&\frac{1}{17010 n^10}
(36377 n^{10}-42252 n^9-36535 n^8+72492 n^7-18685 n^6+2772 n^5+12259 n^4 \notag \\
&~&-54012 n^3+13288 n^2+21000 n-6704) \notag\eea
\be C^{-}_{2,0}=C^{-}_{2,1}=C^{-}_{2,3}=0,\hs{3ex}  C^{-}_{2,2}=\frac{4 \left(n^2-1\right)}{n^2} \notag\ee
\be C^{-}_{2,4}=\frac{2 \left(149 n^6-90 n^5-102 n^4+180 n^3-99 n^2-90 n+52\right)}{45 n^6} \notag \ee

\end{appendix}

%%%%%%%%%%%%%%%%%%%%%%%%%% references %%%%%%%%%%%%%%%%%%%%%%%%%%%%%%%%%%%%%%%%%%%%%%%%%%%%%%%%%%%%%%%%%%%%%%%

  \end{document}